\documentclass{aa}

\usepackage[dvips]{graphicx}
\usepackage{txfonts}
\usepackage{natbib}
\usepackage{color}
\usepackage{MR_AA}

\begin{document}

\title{Self-consistent 2D models of fast-rotating early-type stars}

\subtitle{}

\author{F. Espinosa Lara\inst{1,2}
\and M. Rieutord\inst{1,2}} 

\institute{Universit\'e de Toulouse; UPS-OMP; IRAP; Toulouse, France
\and CNRS; IRAP; 14, avenue Edouard Belin, F-31400 Toulouse, France}

\date{\today}

\abstract{}{This work aims at presenting the first
two-dimensional models of an isolated rapidly rotating star that include
the derivation of the differential rotation and meridional circulation
in a self-consistent way.
}{
We use spectral methods in multidomains, together
with a Newton algorithm to determine the steady state solutions
including differential rotation and meridional circulation for an
isolated non-magnetic, rapidly rotating early-type star. In particular we
devise an asymptotic method for small Ekman numbers (small viscosities) that
removes the Ekman boundary layer and lifts the degeneracy of the
inviscid baroclinic solutions.
}{
For the first time, realistic two-dimensional
models of fast-rotating stars are computed with the actual
baroclinic flows that predict the differential rotation and the
meridional circulation for intermediate-mass and massive stars. These
models nicely compare with available data of some nearby fast-rotating
early-type stars like Ras Alhague ($\alpha$ Oph), Regulus ($\alpha$ Leo),
and Vega ($\alpha$ Lyr). It is shown that baroclinicity drives a differential
rotation with a slow pole, a fast equator, a fast core, and a slow
envelope. The differential rotation is found to increase with mass, with
evolution (here measured by the hydrogen mass fraction in the core), and
with metallicity. The core-envelope interface is found to be a place of
strong shear where mixing will be efficient.
}{
Two-dimensional models offer a new view of fast-rotating stars,
especially of their differential rotation, which turns out to be strong
at the core-envelope interface. They also offer more accurate models for
interpreting the interferometric and spectroscopic data of
early-type stars.}


\maketitle

\section{Introduction}

One of the main challenges of current stellar physics is to
understand the overall dynamics of stars. Indeed, the new
detailed observations of stars that report either on magnetic fields
\cite[][]{petit_etal09,petit_etal10} or on abundance patterns in clusters
\cite[][]{Hunter_etal07,brott_etal11} both crucially depend on the fluid
flows that pervade the interior of stars. Thermal convection has long
been related to the generation of magnetic fields, but the dynamics of
convectively stable radiative regions has been considered for at
least twenty years as the cornerstone in understanding surface
abundances in stars. Moreover, the heart of these large-scale dynamics is
the ubiquitous rotation of stars. It is now well-known that rotation is a
key parameter for stellar dynamos \cite[][]{petit_etal08}, but it is also
crucial to the macroscopic transport through stably stratified radiative
regions.  There, the baroclinic (or other) flows along with some
small-scale turbulence determine the so-called rotational mixing that is
now a key process in interpreting stellar abundances \cite[][]{MM00}. As a result,
progress toward understanding stars and interpreting the
new data sets, that come from spectropolarimetry (magnetic fields),
spectroscopic surveys, interferometry, and seismology requires a precise
account of the distribution and evolution of angular momentum in stars.

The most common implementation of rotation and rotational mixing in
stellar evolution codes is the prescription of \cite{zahn92} and its
subsequent improvements \cite[][]{TZ97,MZ98}. This modelling is
designed for one-dimensional codes. It relies on an averaging of the
differential rotation, which becomes purely radial (shellular), and of
the meridional transport, which, thanks to some natural hypothesis on
the turbulence in a radiative zone, reduces to an effective diffusion
for chemicals or a simple advection and diffusion for angular momentum
\cite[see][]{chaboyer_Z92,zahn92}.  As underlined by \cite{zahn92},
this model rests on the conjecture that all deviations from equilibrium
can be represented by a single spherical harmonic, namely $Y_2(\cth)$
also standing for the first effect of centrifugal distortion. The
formal expansion, including the whole spherical harmonics series, has been
derived in \cite{MZ04,MZ05} and is currently included in the
one-dimensional code STAREVOL
\cite[][]{decressin_etal09}, but deviations from sphericity must
remain small. Such models are therefore expected to be valid for slow
rotation since fast rotation generates more complicated flows and a
significant distortion of the star. However, exploration of the various
stellar situations requires investigating cases where slow rotation is
not possible \cite[e.g.][]{ekstrom_etal12}. Clearly, we lack a
more detailed view of the large-scale and long-time-scale dynamics of
stars, in the first place to put bounds on the validity of 1D-models.

A closer view at the stellar (fluid) dynamics clearly requires
laying aside the spherically symmetric models. Modelling dynamics needs
more than one dimension, thus a general use of two-dimensional models is the
obvious next step in stellar modelling.

However, jumping into the multidimensional modelling of stars is not a sinecure.
Attempts have followed one another since the first polytropes of
\cite{james64} \cite[see also][for a review of the historical
landmarks of 2D-models]{R06c}, but presently no code can represent stellar
evolution in two (or more) space dimensions, including self-consistently
the large-scale dynamics. Most of the present models impose an ad hoc
law of internal rotation. The reason is that
computing the baroclinic flows (or others), together with the stellar
structure, is difficult numerically because equations are stiff.

These foreseeable difficulties motivated the dedicated studies of
\cite{R06} and \cite{ELR07}. In \cite{R06} the specific features of baroclinic
flows were investigated through a simplified model using the
Boussinesq approximation. The next step \cite[][]{ELR07} has been to
compute a completely radiative star enclosed in a spherical
container. With this configuration we could self-consistently compute 
the flows and the hydrostatic structure of the interior of a fast-rotating
 star. Although this computation has already given a good idea of the
differential rotation and the meridional circulation that may be
expected in radiative zones of rotating stars, the spherical shape of
the container enclosing the star is not appropriate to give the surface
characteristics of such stars, which are important from the point of
view of observations. In addition the computation remained limited to
the interior of the star since polar pressure was always more than 10$^{-5}$
times the central one (realistic values are of the order of 10$^{-12}$ or
less). The natural next step that we present here therefore demanded 
serious effort to overcome the simplification of the previous models. We
required that the geometry of the coordinates follows the surface of the
star so that boundary conditions could be correctly implemented and that
solutions extend up to the true surface of the star as given by an optical
depth condition.

While heading to that goal, it turned out that computating velocity
fields was more difficult than ever, essentially because of the huge
variations in density along the radius of the star (typically more than eight
orders of magnitude).

The present paper aims at reporting these latest advances in the
building of the two-dimensional models of rapidly rotating stars. In
the following section we summarize our strategy for constructing
these models and report on the tests that were made to validate them. In
section 3, we discuss the problem of the velocity fields and present the
boundary layer analysis necessary for using asymptotic solutions.
The next section presents some preliminary results including a
comparison with observational data and a first survey of the dependence
of differential rotation on the general parameters of intermediate-mass
stars. Some conclusions end the paper.

\bigskip
\bigskip
\section{Construction of 2D-models}

\subsection{Hypothesis}

As in \cite{ELR07}, we restrict our models to those describing isolated
rotating stars that are in a steady state. This idealization, which
neglects both chemical evolution and any mass loss, is of course
a first approach to the much more complex reality. In addition, we
consider the star strictly axisymmetric. Breaking of this symmetry
comes from turbulence and magnetic fields. Turbulence is not a problem
in principle, since we are interested in long-term averages. Effects of
turbulence should be quasi-steady and axisymmetric (although their correct
modelling is still a major issue). Magnetic fields, on the contrary, cannot
be ignored a priori. Non-axisymmetric, steady magnetic fields are known to
exist in some rotating stars (i.e. roAp stars for instance). However, as
progress from simple to complicate commands it, we first ignore magnetic
fields. This may not be so bad for some A stars like Vega, whose surface
magnetic fields are rather weak \cite[][]{petit_etal10}.

\subsection{Equations}

The partial differential equations that determine the steady state of a lonely
rotating star are the following:

\greq
     \Delta\phi = 4\pi G\rho \\
     \rho T \vv\cdot\na S = -\Div\vF + \eps_*\\
     \rho \vv\cdot\na\vv = -\na P -\rho\na\phi+\vF_v\\
     \Div(\rho\vv) = 0.
\egreqn{basiceq}
There, one recognizes Poisson's equation ($\phi$ is the gravitational
potential, $\rho$ the density, and $G$ the gravitation constant),
the equation of entropy $S$ ($T$ is the temperature, $\vv$ the velocity,
$\vF$ the heat flux, and $\eps_*$ the nuclear heat sources), the momentum
equation written here in an inertial frame
($P$ is the pressure and $\vF_v$ the viscous force), and lastly the
equation of mass conservation.

These differential equations need to be completed by constitutive laws. We
write the expression of the heat flux as

\beq
\vF = -\khi_r\na T -\frac{\khi_{\rm turb}}{{\cal R}_M}T\na S
\eeq
where $\khi_r$ is the radiative conductivity and $\khi_{\rm turb}$ a
turbulent heat conductivity. Here, ${\cal R}_M={\cal R}/{\cal M}$ where ${\cal
R}$ is the ideal gas constant and ${\cal M}$ the mean molecular mass of
the fluid. We hypothesize that turbulent convection is equivalent
to the diffusion of entropy. As for the viscous force, the similar level
of modelling turbulence leads us to use a turbulent viscosity. We
therefore assume that

\beqan 
\vF_v = \mu\vec{\calF}_\mu(\vv) &=&
\displaystyle\mu\lc\Delta\vv+\frac{1}{3}\na\left(\na\cdot\vv\right)
+2\left(\na\ln\mu\cdot\na\right)\vv\right.\nonumber \\
&&\displaystyle
\quad\left.+\na\ln\mu\times(\na\times\vv)
-\frac{2}{3}\left(\na\cdot\vv\right)\na\ln\mu\rc \;,
\eeqan{fvisc}
where $\mu$ is the dynamical viscosity of the gas.

The equations are also completed by the relations 

\greq
     P\equiv P(\rho,T)\\
     \kappa \equiv \kappa(\rho,T)\\
     \eps_* \equiv \eps_*(\rho,T)\; ,
\egreq
which respectively give the equation of state, the opacity, and
the nuclear heat generation. We recall that in radiative diffusive
equilibrium, heat conductivity is related to opacity by

\beq \khi_r = \frac{16\sigma T^3}{3\kappa\rho}, \eeq
where $\sigma$ is the Stefan-Boltzmann constant. For the energy generation
we use an analytical formula that represents hydrogen burning either in
CNO or pp-chains, namely:

\beq 
\eps_*(\rho,T,X,Z) = \eps_0(X,Z)\rho^2T^{-2/3}\exp\lp A/T^{1/3}\rp 
\eeqn{enuc}
as in \cite{ELR07}. It is completed by the use of OPAL tables for
computing the opacity and deriving the density
from the equation of state ($X=0.7$ and $Z=0.02$ with solar
composition of \citealt{GN93}).

\subsection{Boundary conditions}

The foregoing partial differential equations require boundary conditions
for the solutions to be fully determined. These apply at the star centre
and surface. The conditions at the centre are that the solutions should
simply be regular. Surface boundary conditions are more involved.

The first outer boundary condition is  on the gravitational potential,
which should match that of a field in the vacuum, vanishing at infinity. A
way of solving this difficulty is to use the analytical solution
of Poisson's equation using the Green function. This is the so-called
self-consistent field method developed by \cite{OM68} and used in 2D
models of \cite{JMS05}. We prefer another technique that consists in
computing this field in a surrounding empty domain bounded by a sphere
where simple boundary conditions can be applied \cite[see][]{REL12}.

The next boundary conditions are those to be imposed on the velocity
field. For an isolated star it is natural to use stress-free conditions:

\[ \vv\cdot \vn =0 \quad {\rm and}\quad ([\sigma]\vn)\wedge\vn =\vzero \]
where $[\sigma]$ is the stress tensor. The fluid is assumed to not flow
outside the star and to feel no horizontal stress ($\vn$ is the outer
normal of the star).

Finally, the temperature field requires its boundary conditions. We
assume that the star radiates locally as a black body so impose

\beq -\khi_r\vn\cdot\na T = \sigma T^4\;. \eeqn{bctemp}

Both of the velocity and temperature boundary conditions need to be
applied on the surface of the star, which is ill-defined. It depends on
an arbitrary criterion like the value of an optical depth (most commonly).

In non-rotating models, surface pressure is commonly estimated by

\begin{equation}
p_s=\frac{g_s}{\kappa_s}\tau_s
\end{equation}
where subscript `{\it s}' indicates surface quantities, and $\tau_s$ is the
optical depth at the photosphere usually set to 2/3, but $\tau_s=1$ is
also used \cite[][]{Morel97}.

Similarly, we define the bounding surface of the star, i.e. the surface where
boundary conditions are taken, as the isobar where the pressure is

\beq P_s=\tau_s\frac{g_{\rm pole}}{\kappa_{\rm pole}},\eeq
On this isobar, $T=T_{\rm eff}$ only at the pole. On this surface
\eq{bctemp} is therefore not met except at the pole. In fact, to determine the
temperature field, we just need to give the temperature on this surface
as a function of latitude. For that, we assume that the part of the star
that rests above this bounding surface can be described by a polytrope
of index $n$. Thus the pressure verifies

\beq p=P_s(1-\xi)^{n+1}\eeq
where $\xi$ is a non-dimensional variable measuring the height above the
bounding surface. Because this ``atmosphere" is polytropic, we have

\beq T=T_b(1-\xi)\;.\eeq
The top of the atmosphere is the true surface of the star as defined by
a given optical depth $\tau_s$. There the pressure is 

\[ p=\tau_s\frac{g_{\rm eff}}{\kappa}\;,\]
which depends on the co-latitude $\theta$. On the pole, obviously
$p=P_s$. The height of the atmosphere is given by

\[ 1-\xi(\theta) = \lc\frac{\tau_sg_{\rm eff}}{P_s\kappa}\rc^{1/(n+1)} =
\lc\frac{g_{\rm eff}}{g_{\rm pole}}\frac{\kappa_{\rm pole}}{\kappa}\rc^{1/(n+1)}\;, \]
however, on the true surface $T=T_{\rm eff}$, therefore

\[ T_{\rm eff}= T_b \lc\frac{g_{\rm eff}}{g_{\rm pole}}\frac{\kappa_{\rm
pole}}{\kappa}\rc^{1/(n+1)}\;.\]
We therefore find the temperature profile along the bounding surface

\beq T_b(\theta) = \lp\frac{g_{\rm pole}}{g_{\rm eff(\theta)}}\frac{\kappa(\theta)}{\kappa_{\rm
pole}}\rp^{1/(n+1)}\lp\frac{-\khi_r\vn\cdot\na T}{\sigma}\rp^{1/4}\;.
\eeqn{tb}
Finally, on the bounding surface we only need to set

\beq T=T_b(\theta)\;, \eeq
which is quite simple to implement. The polytropic index may be chosen
from a power-law fit of the opacity in the temperature/density range
spanned by the atmosphere. We indeed know that if $\kappa\propto
\rho^\ell T^{-m}$, then $n=(m+3)/(\ell+1)$. This is not a rigorous
description of these layers, which are also baroclinic, but it should be
a reasonable approximation to determine the fundamental parameters of
the star. Nevertheless, if required, a more realistic atmospheric model
can be used to determine $T_b(\theta)$. In the following calculations
we use $\tau_s=1$ and $n=3$. Finally, we also note that the true
equatorial radius is slightly larger than the equatorial radius of the
bounding surface, but the actual difference is always less than 10$^{-3}$ and
is therefore neglected.

\subsection{Angular momentum condition}

Even with the foregoing boundary conditions, the problem is not fully
constrained because the total angular momentum is not specified. To
complete the formulation of the problem we may either enforce the total
angular momentum $L$ or specify the surface equatorial velocity of the star
$V_{\rm eq}$. In the first case we may write

\[ 
\intvol r\sth\rho u_\varphi \,dV = L
\]
or
\[ 
v_\varphi(r=R,\theta=\pi/2) = V_{\rm eq}
\]
in the second case.

\subsection{Numerical method}

We now briefly summarize on the numerical method we use. We refer the reader to
\cite{REL12} for a more detailed account.
To apply (more) easily boundary conditions we use coordinates
that follow the spheroidal shape of the star. The stellar domain is
divided into subdomains.  Inside such a subdomain the relation between
spherical coordinates $(r,\theta,\varphi)$ and our new spheroidal
coordinates ($\zeta,\theta',\varphi'$) is simply

\greq
  r=a_i\zeta+A_i(\zeta)[R_{i+1}(\theta)-a_i\eta_{i+1}]+B_i(\zeta)[R_i(\theta)-a_i\eta_i]\\
\theta=\theta'\\
 \varphi=\varphi'
\egreq
in the i-th domain such that $\eta_i\leq\zeta\leq\eta_{i+1}$. In this
expression, $R_i(\theta)$ is the radius of the i-th bounding surface,
and $A_i(\zeta), B_i(\zeta)$ are third-order polynomials chosen so that,
with the choice of the constants $a_i$ the mapping is continuous and
derivable \cite[see][for more details]{REL12,BGM98}. These
coordinates are the natural coordinates associated with the shape of
the star, but unfortunately they are not orthogonal \cite[see][]{REL12}.

The spatial discretization of the equations is done by using 
spectral methods so as to minimize the size of matrices that arise.
The fields are expanded into a series of spherical harmonics for
their horizontal dependence on the $\theta'$ coordinate, and the radial
functions are sampled on the Gauss-Lobatto collocation nodes, which are
associated with Chebyshev polynomials.

The nonlinear equations of the stellar structure are solved
iteratively. Two methods have been tested: the first is a relaxation
method also known as the fixed-point scheme inspired by \cite{BGM98}
and \cite{ELR07}, while the second is the well-known Newton method. The
fixed-point scheme can be easily coded, while Newton's scheme coding is
rather involved. However, it turns out that the first scheme is really
efficient in simple configurations (polytropes or slowly rotating stars),
while Newton's scheme converges in most cases rather rapidly (typically in
ten iterations when a non-rotating model is given as a starting point).
We therefore kept Newton's method.

\subsection{Calibrations}

The solutions can be tested in various ways. First, the internal accuracy
is easily accessible with spectral methods since the spectral expansion
readily shows whether the spatial resolution is appropriate or not. A slight
change in the initial guess of the solutions allows us to test the
influence of the round-off errors.

When the iterations have converged, we test the solution in two ways:
first with the virial theorem, which relates integral quantities derived
from the momentum equation and, second, with an energy test that also
checks integral quantities but from the energy equation. Details of
these tests may be found in \cite{ELR07} or \cite{REL12}. The virial
test is usually passed with an accuracy better than a few $10^{-12}$,
while the energy test may show errors up to $10^{-5}$. This much looser
behaviour of the energy side comes from the use of the opacity tables,
which generate some noise (either of numerical or physical origin).

\begin{table}
\begin{center}
\begin{tabular}{ccccc}
\hline
Mass   &$\delta R/R$ & $\delta L/L$ &$\delta \rho_c/\rho_c$ &$\delta T_c/T_c$  \\
3      &$10^{-3}   $ & $3\times10^{-3}$&$5\times10^{-3}$ &$8\times10^{-3}$ \\
7      &$6\times10^{-3}$ & $3\times10^{-2}$&$5\times10^{-2}$ &$3\times10^{-4}$ \\
\hline
\end{tabular}
\end{center}
\caption[]{Comparison of the results between the
ESTER code without rotation and the one-dimensional code TGEC.}
\label{compar}
\end{table}

\begin{table}
\begin{center}
\begin{tabular}{ccc}
\hline
                         & \cite{deupree_etal12} & ESTER \\
R$_{\rm eq}$ (R$_\odot$) &   3.006               & 2.899 \\
L (L$_\odot$)            &   33.45               & 33.00 \\
T$_{\rm eq}$ (K)         &   7735                & 7798  \\
T$_{\rm pol}$ (K)        &   9135                & 9229  \\
V$_{\rm eq}$ (km/s)      &   236                 & 238   \\
\hline
\end{tabular}
\end{center}
\caption[]{Comparison of the results between the 2D models of
\cite{deupree_etal12} and ours for $\alpha$ Ophiuchi. The assumed
parameters are the mass, M=2.25~\msun, the chemical composition ($X_{\rm
core}=0.25, X_{\rm env}=0.7, Z=0.02$) and the ratio $R_{\rm pol}/R_{\rm
eq}=0.838$. The ESTER model has been computed with 32 spherical
harmonics and 330 radial grid points spread on 8 spectral subdomains.}
\label{deupreeandus}
\end{table}

Finally, we compared the solutions given by our code at zero rotation
with solutions given by other codes using completely different
numerical methods. We show in Table~\ref{compar} one such comparison
with the one-dimensional code TGEC (Toulouse-Geneva Evolution Code,
see \citealt{hbh08}, \citealt{theado_etal12}).  Similar results are
obtained with the CESAM code \cite[][]{Morel97}. The small discrepancies
between our code and the foregoing codes comes from the absence of
evolution in our code, which allows us to use an analytic formula for
the heating by nuclear reactions, i.e. \eq{enuc}. We also find similar
agreement with the values published by \cite{deupree_etal12}, although
with a slight (4\%) discrepancy on the equatorial radius (see
Table~\ref{deupreeandus}).

\section{The velocity field}

\subsection{General preliminaries}

The velocity field deserves special attention because it is the solution of
stiff partial differential equations. In an isolated steady, rotating
star, flows come from Reynolds stresses and baroclinic torques. The
baroclinic torques are created by the misalignment of pressure and
density gradients
and have to be balanced by the gradient of angular velocity, so that the star
rotates differentially \cite[e.g.][]{R05}.

In an inertial frame the velocity field is governed by

\greq
     \rho (\vv\cdot\na)\vv = -\na P -\rho\na\phi+\vF_v\\
     \Div(\rho\vv) = 0.
\egreqn{mominer}
If we scale the velocity setting $\vv=2\Omega_0R\vw$ and use the
equatorial radius $R$ as the
length scale ($\Omega_0$ is an angular velocity to be specified later),
we can rewrite these two equations in their non-dimensional form:

\greq
     \rho (\vw\cdot\na)\vw = -\na p -\rho\na\phi+E\mu\calF_\mu(\vw)\\
     \Div(\rho\vw) = 0
\egreqn{equ}
where the pressure and the potential have been scaled appropriately. We
thus introduced the Ekman number

\begin{equation}
\label{eq:ekman}
E=\frac{\mu_0}{\rho_0\Omega_0 R^2}
\end{equation}
where $\mu_0$ is a typical value of the dynamical viscosity and $\rho_0$
is the scale of density.

The axisymmetry of the solution suggests the decomposition of the velocity
field into two contributions: the differential rotation $\Omega(r,\theta)$
and the meridional circulation $\vu(r,\theta)$. We thus write

\begin{equation}
\label{eq:vdecomp}
\vec w(r,\theta)=\vec u(r,\theta)+\vec\Omega(r,\theta)\times\vec r
=\vec u+\Omega s\vec{\hat\varphi} \;.
\end{equation}
Here, $(r,\theta,\varphi)$ are the usual spherical coordinates and
\mbox{$s=r\sin\theta$} is the distance to the rotation axis.

Following (\ref{eq:vdecomp}), the momentum equation can be split into
an azimuthal part

\begin{equation}
\label{eq:mom_az}
\rho\vu\cdot\na\Omega+2\Omega\rho\frac{\vec{\hat s}\cdot\vu}{s}=
E\mu\left(\Delta\Omega+\frac{2}{s}\frac{\partial\Omega}{\partial s}
+\na\Omega\cdot\na\ln\mu\right)
\end{equation}
and a meridional part

\begin{equation}
\label{eq:mom_mer}
\rho(\vu\cdot\na)\vu-\rho s\Omega^2\vec{\hat s}=
-\na p-\rho\na\phi+E\mu\vec{\mathcal{F}}_\mu(\vu) \;.
\end{equation}
Taking the curl of this expression (divided by $\rho$) 
we obtain the vorticity equation associated with the 
meridional component of the velocity field

\begin{equation}
\label{eq:vorticity}
\na\times(\vo\times\vu)-s\frac{\partial\Omega^2}{\partial z}\vec{\hat\varphi}=
\frac{\na \rho\times\na p}{\rho^2}+E\frac{\mu}{\rho}\Delta\vo+E\vec{\calM}_\mu(\vu)
\end{equation}
with

\begin{equation}
\begin{array}{rl}
\vec{\calM}_\mu(\vu)=&
\displaystyle
\na\left(\frac{\mu}{\rho}\right)\times\vec{\calF}_\mu(\vu)
+\frac{\mu}{\rho}\na\times\lc\frac{}{}2\left(\na\ln\mu\cdot\na\right)\vu\right.\\
&\displaystyle
\quad\left.+\na\ln\mu\times(\na\times\vu)
-\frac{2}{3}\left(\na\cdot\vu\right)\na\ln\mu\rc
\end{array}
\end{equation}
where $\vo=\na\times\vu$ is the vorticity of the meridional
circulation. Only the $\varphi$-component of this equation
is non-zero.

Equation (\ref{eq:mom_az}) can be easily transformed into an equation for the 
transport of angular momentum

\begin{equation}
\label{eq:angular_mom}
\na\cdot{(\rho s^2\Omega\vu)}=E\na\cdot (\mu s^2\na\Omega)
\end{equation}
establishing that, for the fluid to be in a stationary state, the
transport of angular momentum by the viscous force must be balanced by
the transport induced by the meridional circulation.

\subsection{Viscosity in stars}

\begin{figure}
\resizebox{\hsize}{!}{\includegraphics{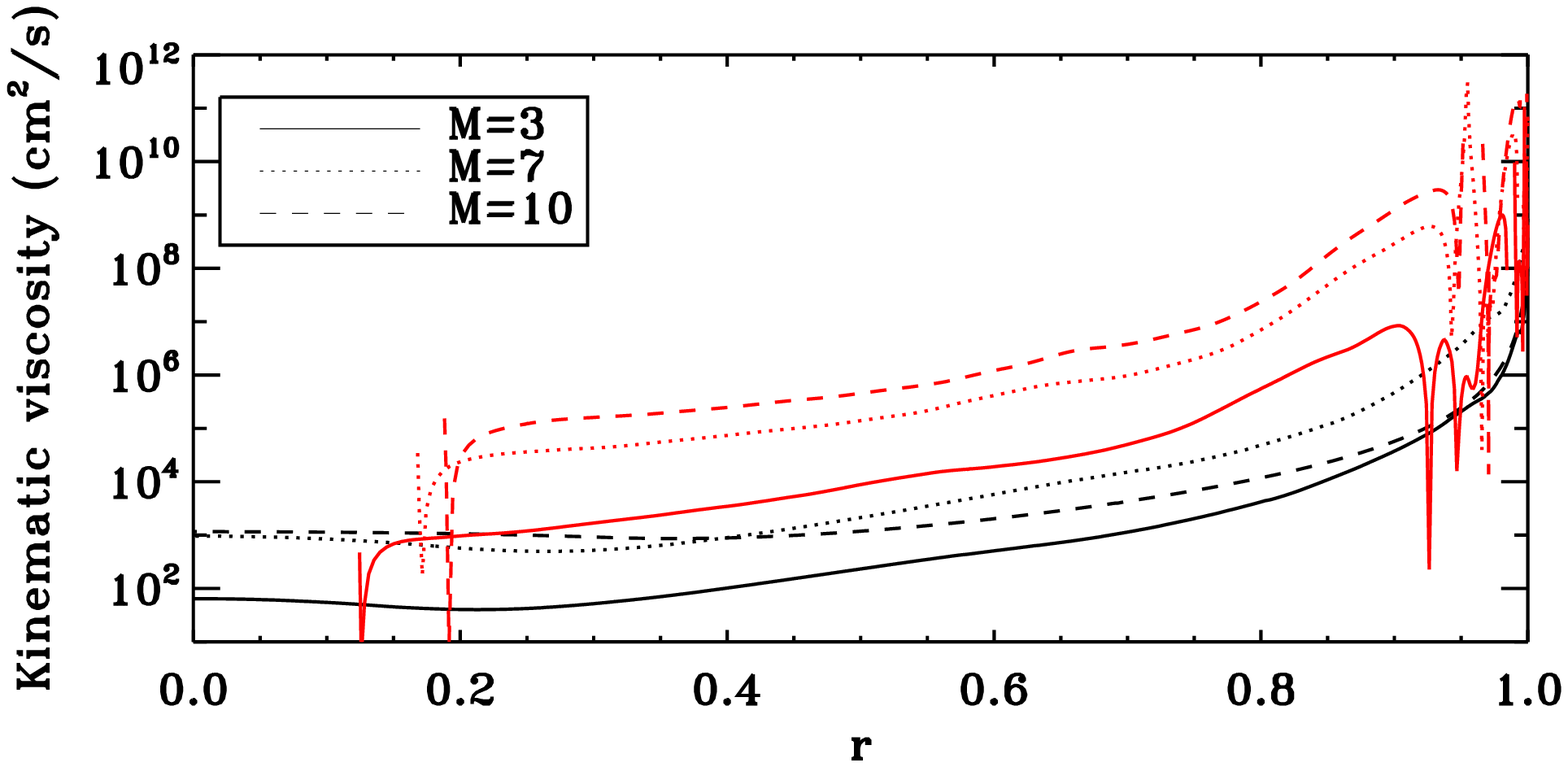}}
\resizebox{\hsize}{!}{\includegraphics{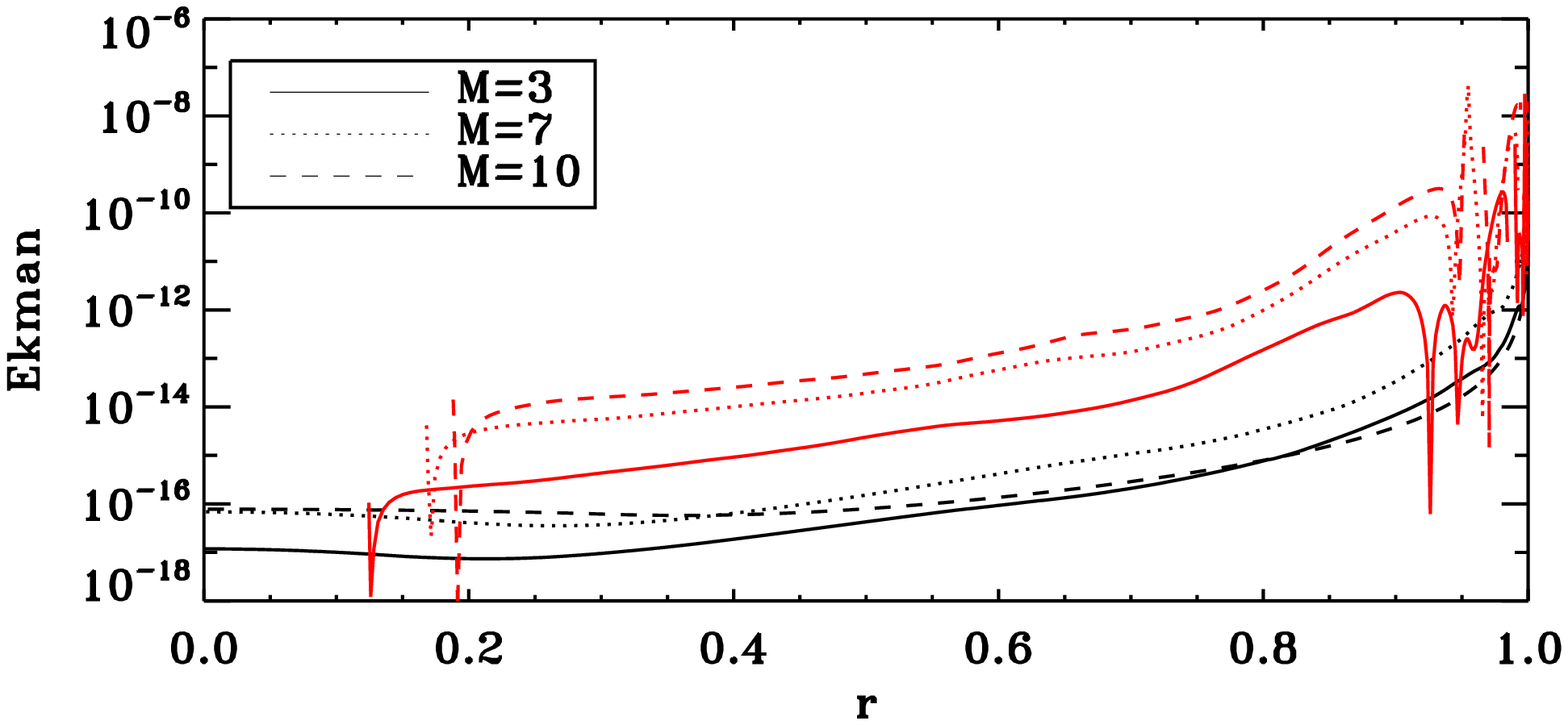}}
\caption{Top: Profiles of the radiative (black) and turbulent (red) viscosity 
as a function of the normalized radius, for three stellar models
of ZAMS stars of 3~\msun, 7~\msun\  and 10~\msun. Bottom: the associated
profiles of the Ekman number $\nu/2\Omega R^2$ based on the radius of
the star and a rotation period of 12~hrs.}
\label{ekman}
\end{figure}

For the stars we are considering that have no convective envelope, the
physical conditions are such that radiative viscosity largely dominates
that of collisional origin. In Fig.~\ref{ekman}, we plot the values
of this viscosity along with the associated Ekman number based on the
radius of the star and an angular velocity corresponding to a rotation
period of 12 hours. This figure shows that the radiative viscosity leads
to Ekman numbers always less than $10^{-10}$.

Because of these low values, some turbulence may develop as a result of
the shear of the differential rotation. \cite{zahn92} proposes that the
induced turbulent vertical viscosity reads as

\[ \mu_t = \rho\frac{\RI_c\kappa}{3}\lp\frac{s}{\calN}\dns{\Omega}\rp^2
\]
where $\RI_c$ is the critical Richardson number. Assuming that
$\RI_c\sim1/4$, we show in Fig.~\ref{ekman} the values of the turbulent
Ekman number, with the same scalings as above, computed from the
differential rotation of a 2D-model. Although it is a few orders of magnitude
larger, the turbulent Ekman number of fast-rotating stars is still very
small compared to unity (always less than $10^{-7}$).

\subsection{Inviscid solution and the occurrence of a boundary layer}

The extremely low values of the stellar Ekman numbers lead us to
look for solutions to the velocity equations that are valid in the asymptotic
regime \mbox{$E\rightarrow0$}.  In this regime, the viscous force is not
able to maintain a solid body rotation. It therefore carries angular
momentum. In a steady state, this flux is balanced by the meridional flow
according to (\ref{eq:angular_mom}). This expression gives the
order of magnitude of this meridional flow. Thus, we have

\begin{equation}
\|\vu\|\sim O(E)\;.
\label{odu}
\end{equation}
Using this result, the meridional flow can be decoupled from the rest
of the equations. Indeed, all the terms depending on $\vu$ in the other
components of the momentum equation are $O(E^2)$ and can be neglected.
In particular, the meridional projection will become

\begin{equation}
\rho s\Omega^2\vec{\hat s}=\na p+\rho\na\phi
\label{baroc}
\end{equation}
and the vorticity equation

\begin{equation}
\label{eq:vort_inv}
s\frac{\partial\Omega^2}{\partial z}=\vec{\hat\varphi}\cdot\frac{\na p\times\na\rho}{\rho^2} \;.
\end{equation}
In principle, (\ref{eq:vort_inv}) can be used to calculate the
rotation profile for a given configuration. Unfortunately, without
adequate boundary conditions, the rotation profile is undetermined by a
function of $s$; that is if $\Omega(\vec r)$ is a solution, $\Omega'^2(\vec
r)=\Omega^2(\vec r)+F(s)$ is also a solution.

This degeneracy is removed by viscosity. Indeed, the free surface of
the star is assumed to support no stress, so the angular velocity
should verify

\begin{equation}
\vec{\hat n}\cdot\na\Omega=0 \;,
\end{equation}
where $\vec{\hat n}$ is the unit vector normal to the surface. However,
this condition is incompatible with (\ref{eq:vort_inv}). This is a
standard problem that is solved using a boundary layer analysis. The
main idea is to take advantage of the existence of a thin layer in
the vicinity of the boundary where viscosity is not negligible and
where the flow adapts its properties to fulfil the boundary conditions. Doing this, we can
separate the problem into two domains, an interior one where viscosity
can be neglected and a thin boundary layer, dominated by viscosity. In
our case, the thickness of the boundary layer depends on the Ekman
number. It is infinitesimally thin when $E\rightarrow0$, which is used
to simplify the equations, since the vertical gradient of a boundary layer
quantity is much larger than the horizontal one. This procedure has been
successfully used to derive baroclinic flows in a Boussinesq model of
a star \cite[see][]{R06}. We now adapt it to realistic stellar models.

\subsection{Stretched coordinates}
\label{bl_coords}

It is useful to define a new set of stretched coordinates
$(\xi,\tau,\varphi)$ to represent the fluid inside the boundary layer.
The vertical coordinate $\xi$ measures the distance to the surface scaled
by a parameter $\varepsilon$ that represents the thickness of the boundary
layer. The angular coordinate $\tau$ is equal to the colatitude of the
nearest point on the surface, so that $\tau=\theta$ when $\xi=0$.
When choosing that $\xi$ increases outwards, $\xi$ is negative inside the
star. The azimuthal coordinate $\varphi$ coincides with the corresponding
spherical coordinate.

\begin{figure}
\resizebox{\hsize}{!}{\includegraphics{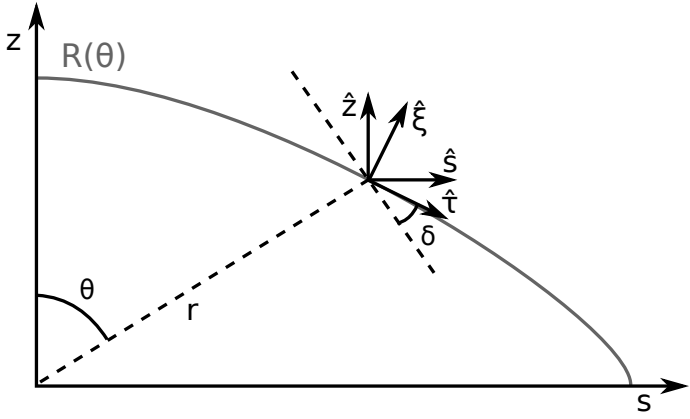}}
\caption{A schematic view of the local basis adapted to the analysis of
the Ekman layer.}
\label{fig:coords}
\end{figure}

The unit vectors $\vec{\hat\xi}$ and $\vec{\hat\tau}$ are 
perpendicular and parallel to the surface, respectively. They can be expressed as

\begin{equation}
\begin{array}{l}
\vec{\hat\xi}=\vec{\hat s}\sin(\tau-\delta)+\vec{\hat z}\cos(\tau-\delta) \\ 
\vec{\hat\tau}=\vec{\hat s}\cos(\tau-\delta)-\vec{\hat z}\sin(\tau-\delta)
\end{array}
\end{equation}
where $\vec{\hat s}$ and $\vec{\hat z}$ are the unit vectors associated
with the cylindrical coordinates, and $\delta$ is defined as

\begin{equation}
\tan\delta(\tau)=\frac{1}{R(\tau)}\frac{\mathrm{d}R}{\mathrm{d}\tau}
\end{equation} 
and accounts for the deformation of the surface due to the centrifugal force.
Figures~\ref{fig:coords} sketches out this new local frame.
The relation between $(\xi,\tau)$ and the cylindrical coordinates $(s,z)$
is given by

\begin{equation}
\begin{array}{l}
s=R(\tau)\sin\tau+\varepsilon\xi\vec{\hat s}\cdot\vec{\hat\xi}=
R(\tau)\sin\tau+\varepsilon\xi\sin(\tau-\delta)\\
z=R(\tau)\cos\tau+\varepsilon\xi\vec{\hat z}\cdot\vec{\hat\xi}=
R(\tau)\cos\tau+\varepsilon\xi\cos(\tau-\delta) \;.
\end{array}
\end{equation}

The coordinates $(\xi,\tau,\varphi)$ form a set of orthogonal coordinates
with scale factors

\begin{equation}
\begin{array}{l}
\displaystyle h_\xi=\left|\frac{\partial\vec r}{\partial\xi}\right|=
\varepsilon \\
\displaystyle h_\tau=\left|\frac{\partial\vec r}{\partial\tau}\right|=
\frac{R(\tau)}{\cos\delta(\tau)}
+\varepsilon\xi\left(1-\frac{\mathrm{d}\delta}{\mathrm{d}\tau}\right) \\
\displaystyle h_\varphi=\left|\frac{\partial\vec r}{\partial\varphi}\right|=
R(\tau)\sin\tau+\varepsilon\xi\sin(\tau-\delta)\;.
\end{array}
\end{equation}
We are interested in a very thin layer below the surface,
thus $\varepsilon\ll1$ and to leading order

\begin{equation}
\begin{array}{l}
\displaystyle h_\xi=\varepsilon \\
\displaystyle h_\tau=\frac{R(\tau)}{\cos\delta(\tau)}\\
\displaystyle h_\varphi=R(\tau)\sin\tau\;.
\end{array}
\end{equation}

\subsection{Scale analysis}

We start the analysis of the boundary layer by deriving the order
of magnitude of the dependent variables. As we have noted before
(Eq.~\ref{odu}), the meridional velocity in the interior inviscid domain
is $u\sim O(E)$, while the rest of the variables are supposed to be
$\sim O(1)$. Since the flow is steady, mass conservation imposes

\begin{equation}
\vec{\hat n}\cdot\vu=0\quad {\rm at}\quad \xi=0\; .
\end{equation}
Using the boundary layer coordinates, it reads
\mbox{$u_\xi=\vec{\hat\xi}\cdot\vu=0$}, so the normal velocity
$u_\xi$ should go from a value $O(E)$ at the base of the boundary
layer to zero. Thus \mbox{$\frac{\partial u_\xi}{\partial\xi}\sim
O(E)$}, since $\xi$ is a stretched variable. Using the new coordinates
$(\xi,\tau,\varphi)$, the continuity equation now reads as (see appendix
\ref{appendix}) 

\begin{equation}
\frac{1}{\varepsilon}\frac{\partial}{\partial\xi}\left(\rho u_\xi\right)
+\frac{\cos\delta}{R}\frac{\partial}{\partial\tau}\left(\rho u_\tau\right)
+\frac{\cos(\tau-\delta)}{R\sin\tau}\rho u_\tau=0 \;,
\end{equation}
from where we see that $u_\tau\sim O(E/\varepsilon)$.

\begin{figure}
\resizebox{\hsize}{!}{\includegraphics{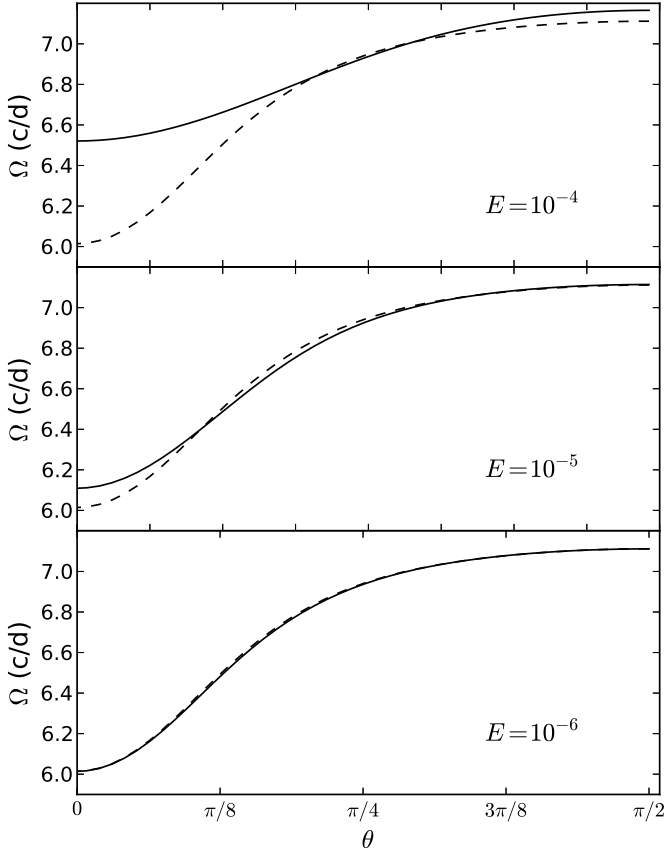}}
\caption{Surface rotation (in cycles per day) as a function of colatitude
for different values of the Ekman number.  The profile approaches the
inviscid solution (dashed line) as the Ekman number decreases.}
\label{surfrot}
\end{figure}

On the surface, the tangential stress should vanish, according to the
stress-free boundary condition. Using the boundary layer coordinates,
this condition reduces to

\begin{equation}
\frac{\partial u_\tau}{\partial\xi}=0 \quad \mbox{and} \quad
\frac{\partial\Omega}{\partial\xi}=0 \;.
\end{equation}
We now go back to the vorticity equation (\ref{eq:vorticity}). If the
vertical gradient of angular velocity vanishes on the surface, the
inviscid vorticity equation \eq{eq:vort_inv} cannot, in general, be
verified. Thus, within the boundary layer, the balance of forces requires
the viscous force whose dominant term $E\frac{\mu}{\rho}\Delta\vec\omega$
should be of order unity.  The vorticity of the meridional velocity
field is

\begin{equation}
\vec\omega=\na\times\vec u=
\left(\frac{1}{\varepsilon}\frac{\partial u_\tau}{\partial\xi}
-\frac{\cos\delta}{R}\frac{\partial u_\xi}{\partial\tau}\right)\vec{\hat\varphi}
\simeq
\frac{1}{\varepsilon}\frac{\partial u_\tau}{\partial\xi}\vec{\hat\varphi}
\end{equation}
and

\begin{equation}
E\frac{\mu}{\rho}\Delta\vec\omega\simeq
E\frac{\mu}{\rho}\frac{1}{\varepsilon^3}\frac{\partial^3 u_\tau}{\partial\xi^3}
\sim O(E^2/\varepsilon^4) \;,
\end{equation}
then the scale of the boundary layer must be

\begin{equation}
\varepsilon=\sqrt{E} \;,
\end{equation}
as expected. Normal and meridional velocities are 
$O(\varepsilon^2)$ and $O(\varepsilon)$, respectively.  It is useful to define the
scaled boundary layer corrections

\begin{equation}
U=\frac{u_\xi-\vec{\hat\xi}\cdot\vec u_\mathrm{int}}{\varepsilon^2} \quad \mbox{and} \quad
V=\frac{u_\tau-\vec{\hat\tau}\cdot\vec u_\mathrm{int}}{\varepsilon} \;,
\end{equation} 
where $\vec u_\mathrm{int}\sim O(\varepsilon^2)$ is the interior
meridional circulation.

Back to the meridional component of the momentum equation
(\ref{eq:mom_mer}), the viscous force is $O(\varepsilon^2)$ in the
direction of $\vec{\hat\xi}$ and $O(\varepsilon)$ in the direction
of $\vec{\hat\tau}$ and, in the limit $\varepsilon\rightarrow0$, the
inviscid equation is still valid within the boundary layer. Then we
suppose that the rest of the dependent variables $p$, $\rho$, $\phi$ are
not perturbed in the boundary layer and can be treated as functions of
$\tau$ alone, because their vertical derivatives $\frac{\partial}{\partial\xi}$
are $O(\varepsilon)$.

The angular velocity $\Omega$ is slightly perturbed within the boundary layer: its vertical
derivative should change from some finite value to zero, thus we set

\begin{equation}
\frac{\partial\Omega}{\partial\xi}=
\left(\frac{\partial\Omega}{\partial\xi}\right)_\mathrm{interior}+\varepsilon\beta \;,
\end{equation} 
where the vertical variation of $\beta$ is not negligible inside the
boundary layer.

\begin{figure*}
\centering
\includegraphics[width=17cm]{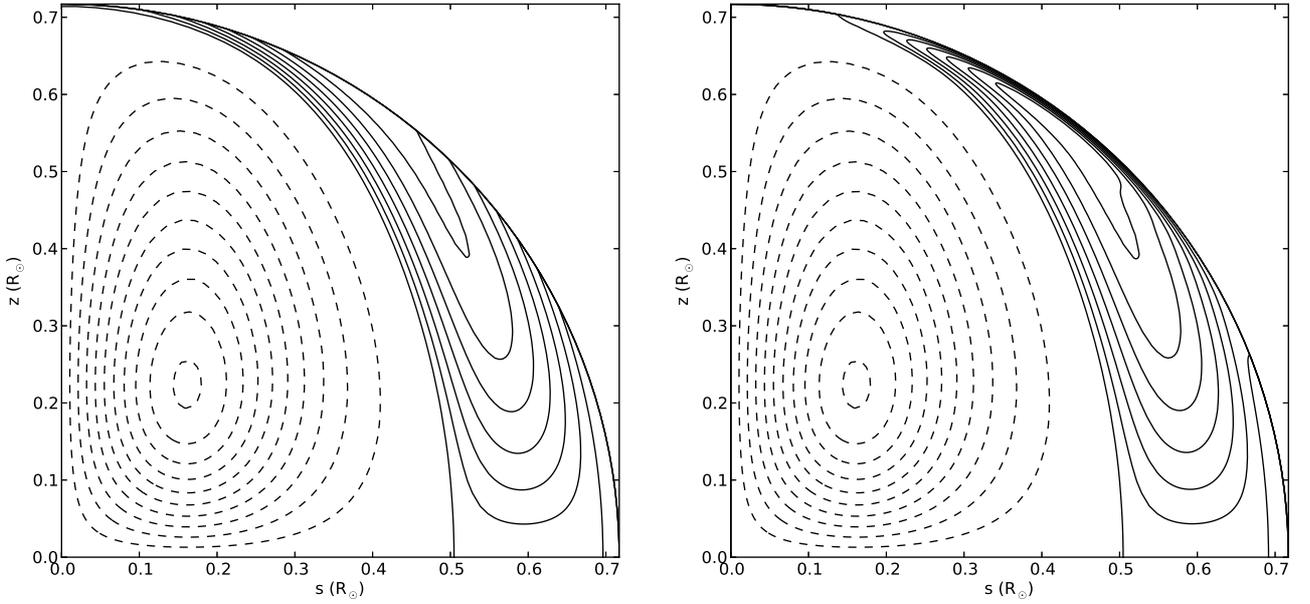}
\caption{Streamlines of the meridional circulation for the inviscid
solution with $E\rightarrow0$ (left) and for the full problem
($E=10^{-7}$) including the viscous force (right).  Solid lines and
dashed lines represent counter-clockwise and clockwise circulation,
respectively. In the viscous solution, the streamlines are
closed inside the boundary layer.  For these calculations, one solar mass
of ideal gas with Kramers' opacities was enclosed in a bounding sphere at
the pole of which the pressure is $10^{-4}$ times the central value
\cite[see][for a description of this model]{ELR07}.}
\label{circ}
\end{figure*}

\subsection{Boundary layer equations}

In short, we have to solve for three variables in the boundary layer:
the boundary layer corrections of the normal and tangential components of
the meridional velocity field $U$ and $V$, and of the vertical gradient
of rotation that we have called $\beta$.  The equations describing
the flow in the boundary layer are obtained from the vorticity equation
(\ref{eq:vorticity}), the conservation of angular momentum \eq{eq:mom_az}
and the continuity equation $\Div(\rho\vu)=0$.  At leading order in
$\varepsilon$ and after subtracting the interior inviscid equations,
these equations read as (see appendix \ref{appendix})

\begin{eqnarray}
\displaystyle
2\Omega\ts\beta\cos(\tau-\delta)+\nu\frac{\partial^3V}{\partial\xi^3}&=&0\;,
\label{eq:bl_vort}\\
\displaystyle
\lp2\Omega+\ts\frac{\partial\Omega}{\partial\ts}\rp V\cos(\tau-\delta)
-\ts\nu\frac{\partial\beta}{\partial\xi}&=&0\;,
\label{eq:bl_angmom}\\
\displaystyle
\frac{\ts}{\cos(\tau-\delta)}\frac{\partial U}{\partial\xi}+
\ts\frac{\partial V}{\partial\ts}+\left(1+\frac{\ts}{\rho}
\frac{\partial\rho}{\partial\ts}\right) V&=&0\;,
\label{eq:bl_cont}
\end{eqnarray}
where we have defined $\ts=R(\tau)\sin\tau$ and $\nu=\mu/\rho$,
$\ts$ is the radial cylindrical coordinate at the surface, and $\nu$ is
the non-dimensional kinematic viscosity.  These equations are completed
with boundary conditions. On the surface ($\xi=0$), the conservation of
mass and the stress-free boundary conditions become

\begin{equation}
\label{eq:bl_bc0}
U=-\frac{1}{E}\vec{\hat\xi}\cdot\vec u_\mathrm{int}\;,
\quad\frac{\partial V}{\partial\xi}=0
\quad\mbox{and}\quad\beta=\beta_0(\tau)\quad\mbox{at}\quad\xi=0 
\end{equation}
where $\beta_0=-\vec{\hat\xi}\cdot(\na\Omega)_\mathrm{int.}$.
The corrections should also vanish in the interior, then

\begin{equation}
U\rightarrow0\;,\quad
V\rightarrow0\quad\mbox{and}\quad\beta\rightarrow0\quad\mbox{when}\quad\xi\rightarrow-\infty\;.
\end{equation}
Combining (\ref{eq:bl_vort}) and (\ref{eq:bl_angmom}), we obtain

\begin{equation}
\frac{\partial^4V}{\partial\xi^4}+\frac{1}{\nu^2}\left(4\Omega^2+
\ts\frac{\partial\Omega^2}{\partial\ts}\right)\cos^2(\tau-\delta)V=0 \;,
\end{equation}
whose solution is

\begin{equation}
V=V_0(\ts)\mathrm{e}^{B\xi}\left[\cos(B\xi)+V_1(\ts)\sin(B\xi)\right]\;,
\end{equation}
with

\begin{equation}
\label{eq:bl_B}
B(\ts)=\sqrt{\frac{\left|\cos(\tau-\delta)\right|}{\nu}}
\left(\Omega^2+\frac{\ts}{4}\frac{\partial\Omega^2}{\partial\ts}\right)^{1/4}\;.
\end{equation}
Here, we have used the fact that $V$ should vanish when
$\xi\rightarrow-\infty$.  Using the boundary condition at the surface
(\ref{eq:bl_bc0}), we see that $V_1=-1$ and therefore

\begin{equation}
\label{eq:bl_V}
V=V_0(\ts)\mathrm{e}^{B\xi}\left[\cos(B\xi)-\sin(B\xi)\right]\;.
\end{equation}
In the previous expressions we have considered that $B$ is real. For that,
$\Omega$ must satisfy the condition

\begin{equation}
\frac{\partial\Omega^2}{\partial\ts}+\frac{4\Omega^2}{\ts}\ge 0 \;,
\end{equation}
which is equivalent to

\begin{equation}
\frac{\partial}{\partial\ts}\left(\ts^2\Omega\right)\ge0 \;.
\end{equation}
In other words, the specific angular momentum must increase with the
distance to the axis of rotation. This is in fact Rayleigh's criterion
for the centrifugal stability of the differential rotation profile.

Using (\ref{eq:bl_vort}) we get $\beta$

\begin{equation}
\label{eq:bl_beta}
\beta=\beta_0(\ts)\mathrm{e}^{B\xi}\cos(B\xi)\;,
\end{equation}
and the condition

\begin{equation}
\label{eq:bl_V0}
V_0(\ts)=\frac{\Omega\ts\cos(\tau-\delta)}{2B^3\nu}\beta_0(\ts)\;.
\end{equation}
Finally, using the continuity equation (\ref{eq:bl_cont}) we obtain the
expression for $U$

\begin{equation}
\label{eq:bl_U}
U=-U_0(\ts)\mathrm{e}^{B\xi}\cos(B\xi)
-U_1(\ts)\xi\mathrm{e}^{B\xi}\left[\cos(B\xi)-\sin(B\xi)\right]\;,
\end{equation}
where

\begin{equation}
\label{eq:bl_U0}
U_0=\frac{\cos(\tau-\delta)}{\ts\rho}\frac{\mathrm{d}}{\mathrm{d}\ts}
\left(\frac{\ts\rho V_0}{B}\right)
\end{equation}
and

\begin{equation}
U_1=\frac{V_0\cos(\tau-\delta)}{B}\frac{\mathrm{d} B}{\mathrm{d}\ts}\;.
\end{equation}

The foregoing expression of the boundary layer correction are those of
the classical Ekman layer adapted for a spheroidal stress-free surface.
Like the Ekman layer, these corrections suffer from the equatorial
singularity. This singularity is associated with the vanishing of
$B(\ts)$ at equator, which is the inverse of the layer's thickness. This is a
quantity of order unity except at the equator where it vanishes
($\tau\tv\pi/2$ and $\delta\tv0$). As was shown by \cite{RS63}, the
boundary layer changes scale in this region which extends  in latitude as 
\od{E^{1/5}}. There, the boundary layer thickness is
\od{E^{2/5}}, which is slightly larger than $E^{1/2}$. The influence of
this singularity tends to zero when the Ekman number vanishes.

\begin{figure*}
\centering
\includegraphics[width=17cm]{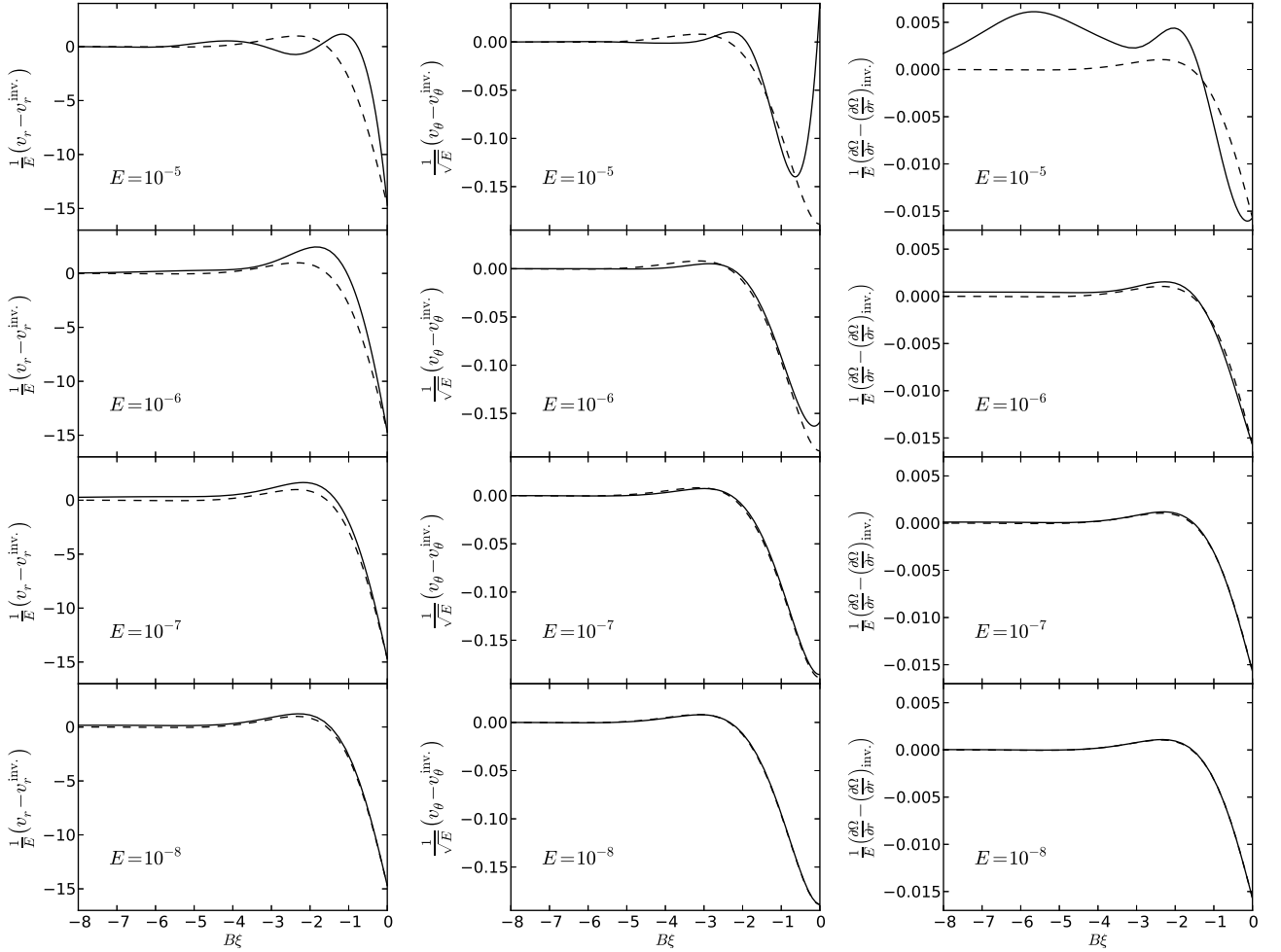}
\caption[]{Boundary layer corrections for the normal velocity (left),
latitudinal velocity (centre), and normal gradient of angular velocity
(right) for different values of the Ekman number $E$. The corrections
are computed at a colatitude $\theta=\pi/4$ and are in normalized
units.  Solid lines show the corrections calculated from the difference
between the full solution and the inviscid model. Dashed lines show the
corrections as predicted by the asymptotic solutions (\ref{eq:bl_V}),
(\ref{eq:bl_beta}), and \eq{eq:bl_U}.}
\label{fig:ekman}
\end{figure*}

\subsection{Surface differential rotation}

The normal velocity in the boundary layer (\ref{eq:bl_U}) should verify
the boundary condition at the surface, as expressed in Eq.~(\ref{eq:bl_bc0}), namely,

\begin{equation}
U(\xi=0)=-U_0=-\frac{1}{E}\vec{\hat\xi}\cdot\vec u_\mathrm{int.}\;.
\end{equation}
We see now that this leads to the appropriate boundary condition on
the $\Omega$  profile at the surface.  As the fluid flow is axisymmetric
and $\na\cdot(\rho\vec u_\mathrm{int})=0$,
the meridional velocity may be described by a stream function $\psi$,
defined as

\begin{equation}
\rho\vec u_\mathrm{int}=\na\times(\psi\vec{\hat\varphi})\;,
\end{equation}
then

\begin{equation}
U_0=\frac{\cos(\tau-\delta)}{\ts\rho}\frac{1}{E}\frac{\partial}{\partial\ts}
\left(\ts\psi\right)\;.
\end{equation}
Using Eq.~(\ref{eq:bl_U0}) we see that, on the surface

\begin{equation}
\label{eq:psi_V0}
\psi(r=R(\theta),\theta)=E\rho\frac{V_0}{B}\;.
\end{equation}
Using Eqs.~(\ref{eq:bl_B}) and (\ref{eq:bl_V0}), we calculate $\frac{V_0}{B}$

\begin{equation}
\label{eq:V0/B}
\frac{V_0}{B}=\frac{\nu\ts}{\left(2\Omega+\ts\frac{\partial\Omega}{\partial\ts}
\right)\cos(\tau-\delta)}\beta_0=
-\frac{\nu\ts^2}{\vec{\hat\tau}\cdot\na(\ts^2\Omega)}
\vec{\hat\xi}\cdot(\na\Omega)_\mathrm{int.}\;.
\end{equation}
Equations (\ref{eq:psi_V0}) and (\ref{eq:V0/B}) lead to the condition for
the interior rotation profile, now expressed using only interior variables

\begin{equation}
\label{eq:bl}
E\mu s^2\vec{\hat\xi}\cdot\na\Omega+\psi\vec{\hat\tau}\cdot\na(s^2\Omega)=0
\qquad\mbox{on the surface}\;.
\end{equation}
This boundary condition is to be imposed to the differential rotation
and lifts the indetermination of the inviscid solution of the
baroclinic flow \eq{eq:vort_inv}. Indeed these solutions
are determined up to an arbitrary function of the radial cylindrical
coordinate $s$. This new condition leads to the differential equation that
determines this arbitrary function. It couples the horizontal derivative
of the angular velocity $\vec{\hat\tau}\cdot\na(s^2\Omega)$ with its
vertical derivative $\vec{\hat\xi}\cdot\na\Omega$. Thus, solving for
this nonlinear boundary condition with the equations of dynamics,
leads directly to a self-consistent differential rotation without having
to solve the Ekman boundary layer. This part of the solution may be
recovered using Eqs. (\ref{eq:bl_V}), (\ref{eq:bl_beta}), and (\ref{eq:bl_U}).

Remarkably enough, this boundary condition is smooth at the equator: the
above-mentioned equatorial singularity is removed with the rest of the
Ekman layer.

\subsection{Numerical validation}

To validate the new boundary condition \eq{eq:bl} to be imposed to the 
inviscid baroclinic flows, we solved numerically the full viscous
equation in the simplified set-up of a star enclosed in a spherical box
as in \cite{ELR07}. As in this previous work we considered a one solar
mass of an ideal gas, whose opacities are given by Kramer's power laws
and which rotates at half the break-up angular velocity. The dynamical
viscosity of the gas is set to a constant $\mu_c$. The interior
is fully radiative, and the polar pressure is set to $2\times10^{-4}$
the central pressure. The latter condition reduces the density
variations and makes the full viscous solution more easily computable.

Because of the imposed constant dynamical viscosity, density variations
induce variations of the kinematic viscosity and thus of the Ekman
number \eq{eq:ekman}.  In Fig.~\ref{surfrot}, we show the convergence of the
surface differential rotation towards the one prescribed by Eq.~\eq{eq:bl}, when
the (central) Ekman number $E_c=\mu/\rho_c2\Omega R^2$ decreases from
$10^{-4}$ to $10^{-6}$. In Fig.~\ref{circ} we show the meridional stream
function in the asymptotic case and for a viscous solution with
$E=10^{-7}$. We note that the two solutions agree nicely.

Finally, we computed the boundary layer corrections for the three
components of the velocity both from the full numerical solution and from
the asymptotic formulae \eq{eq:bl_V}, \eq{eq:bl_beta}, and \eq{eq:bl_U}. As
shown by Fig.~\ref{fig:ekman}, we see that, provided that the Ekman
number is sufficiently low, the two solutions also agree.

\section{Results}

\subsection{The core-envelope interface}

The core-envelope interface (thereafter CEI) is the locus of a complex
dynamics that is important to understand for its consequences on the
lifetime of the stars and the chemical enrichments of their surface
layers.
As a consequence of the baroclinic torque balance \eq{eq:vort_inv}, 
strong density gradient at the CEI (driven for instance by the chemical
evolution of the core) forces a strong angular velocity gradient at this
same place. In our models, which neglect viscous effect in the interior
and any diffusion of elements, this velocity gradient is represented by a
jump in the angular velocity. This jump is controlled by the latitudinal
variations in the pressure along the CEI. Indeed, the CEI is a surface
where the \BVF\ squared changes sign. If there is a density jump there,
$\calN^2$ behaves there as a Dirac, and the baroclinic torque is not
defined. However, since pressure must be continuous, its latitudinal
variations on the CEI just inside and just outside the core must be
the same. From the projection of \eq{baroc} on a tangent vector $\ve_T$
of the CEI, we can therefore derive the following jump condition

\beqan  \rho_{\rm core}(\theta) \lp s\Omega^2_{\rm core}(\theta) \es\cdot\ve_T -
\ve_T\cdot\na\phi_{\rm core}\rp =\nonumber \\
\rho_{\rm env}(\theta) \lp s\Omega^2_{\rm env}(\theta) \es\cdot\ve_T - 
\ve_T\cdot\na\phi_{\rm env}\rp\;.
\eeqan{jcond}
This interface conditions shows that if the density is discontinuous,
so is the angular velocity, since $\na \phi$ is continuous.

\begin{figure}
\resizebox{\hsize}{!}{\includegraphics{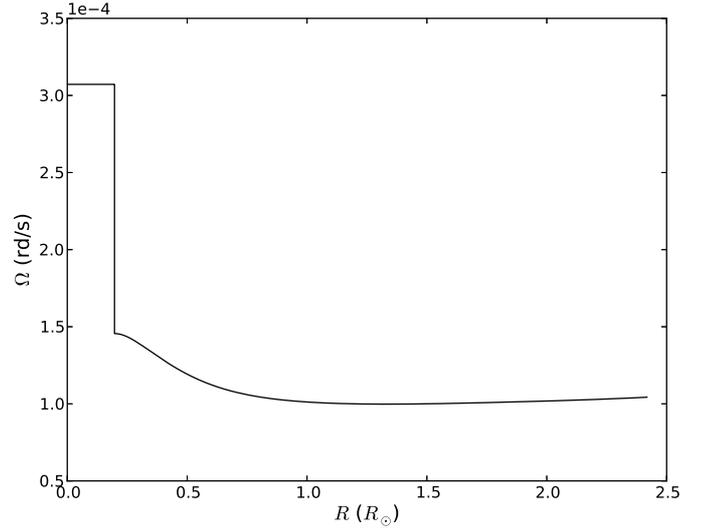}}
\caption{Angular velocity (in rd/s) as a function of the radial distance
(in R$_\odot$) along the rotation axis ($\theta=0$) for the Vega model of
Table~\ref{vega}. The angular velocity discontinuity at the core-envelope
interface is clearly visible.}\label{om_jump}
\end{figure}

Numerics show that if the core is denser than the envelope, it
rotates faster, as illustrated in Fig.~\ref{om_jump}. We note that this
faster rotation of the core and the ensuing velocity gradient builds up
on the thermal time scale of the star. This is indeed the time scale that
controls the baroclinic modes and therefore the action of the baroclinic
torque. This time scale is usually smaller than that of nuclear evolution,
and therefore this velocity jump has time to build up when the core gets
denser as nuclear evolution proceeds.

\begin{figure}
\resizebox{\hsize}{!}{\includegraphics{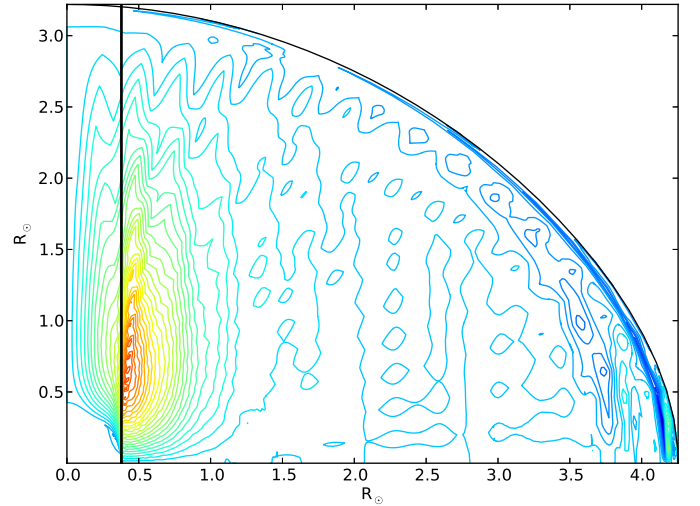}}
\caption{Meridional streamlines from the model of $\alpha$ Leo. The
vertical solid line illustrates the tangent cylinder associated with
the core-envelope interface.} \label{psi_cei}
\end{figure}

The foregoing shear is obviously smoothed by diffusion processes,
in particular by viscosity. However, this smoothing is not simple. The
balance of forces requires the occurrence of a Stewartson layer that develops
on the tangent cylinder of the core-envelope interface as illustrated
in Fig.~\ref{psi_cei}. Such a layer is triggered by any discontinuity
that appears on this interface. As shown in the original analysis of
\cite{stewar66}, this internal layer is composed of nested layers whose
thicknesses scale with some fractional power of the viscosity ($E^{1/4}$,
$E^{2/7}$ and $E^{1/3}$ are the main non-dimensional scales). Such
scalings are not included in our solution, which only retains the balance
between viscous force and meridional advection on a macroscopic scale.

In Fig.~\ref{psi_cei}, we clearly see the meridional flows induced by
the core-envelope interface.  Their spectral expansion on spherical
harmonics is not converged well as a consequence of the hidden
discontinuities. The foregoing remarks aim at underlining that such flows
should be considered as not fully computed.  This is why we do not
comment on the transport they may achieve, although this is one of the
key parts of rotational mixing. The core-envelope shear layer is clearly
reminiscent of the tachocline of low-mass stars, although the mechanisms
at work are different. As for this layer, we should expect some
dynamo effect driven by the shear. Further work is clearly needed.

Fortunately, the differential rotation is computed much better and does
not suffer from the approximate treatment of the Stewartson layer. This
is most probably because the influence of this layer vanishes as the
viscosity vanishes.

\begin{figure*}
\resizebox{\hsize}{!}{\includegraphics{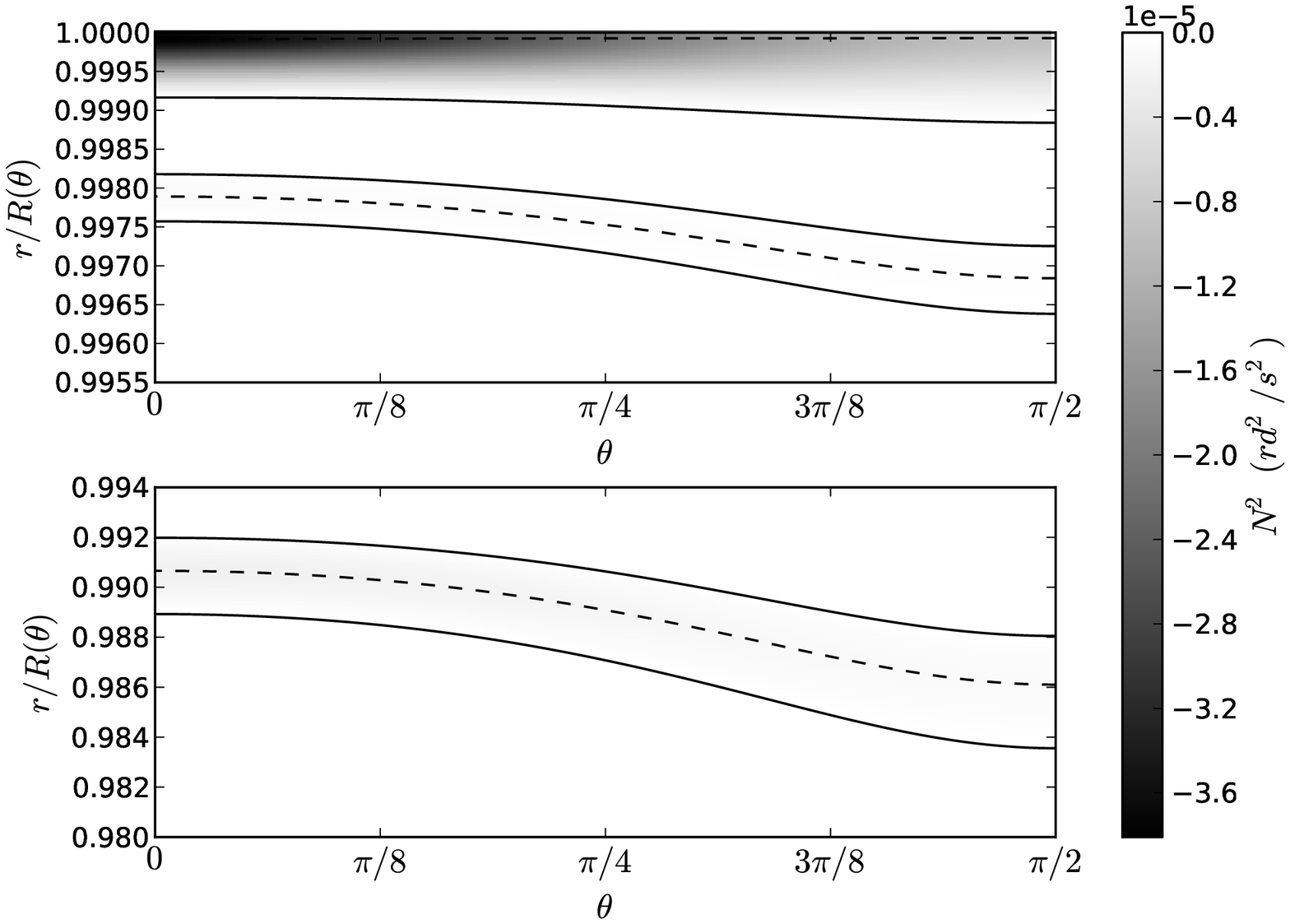}
\includegraphics{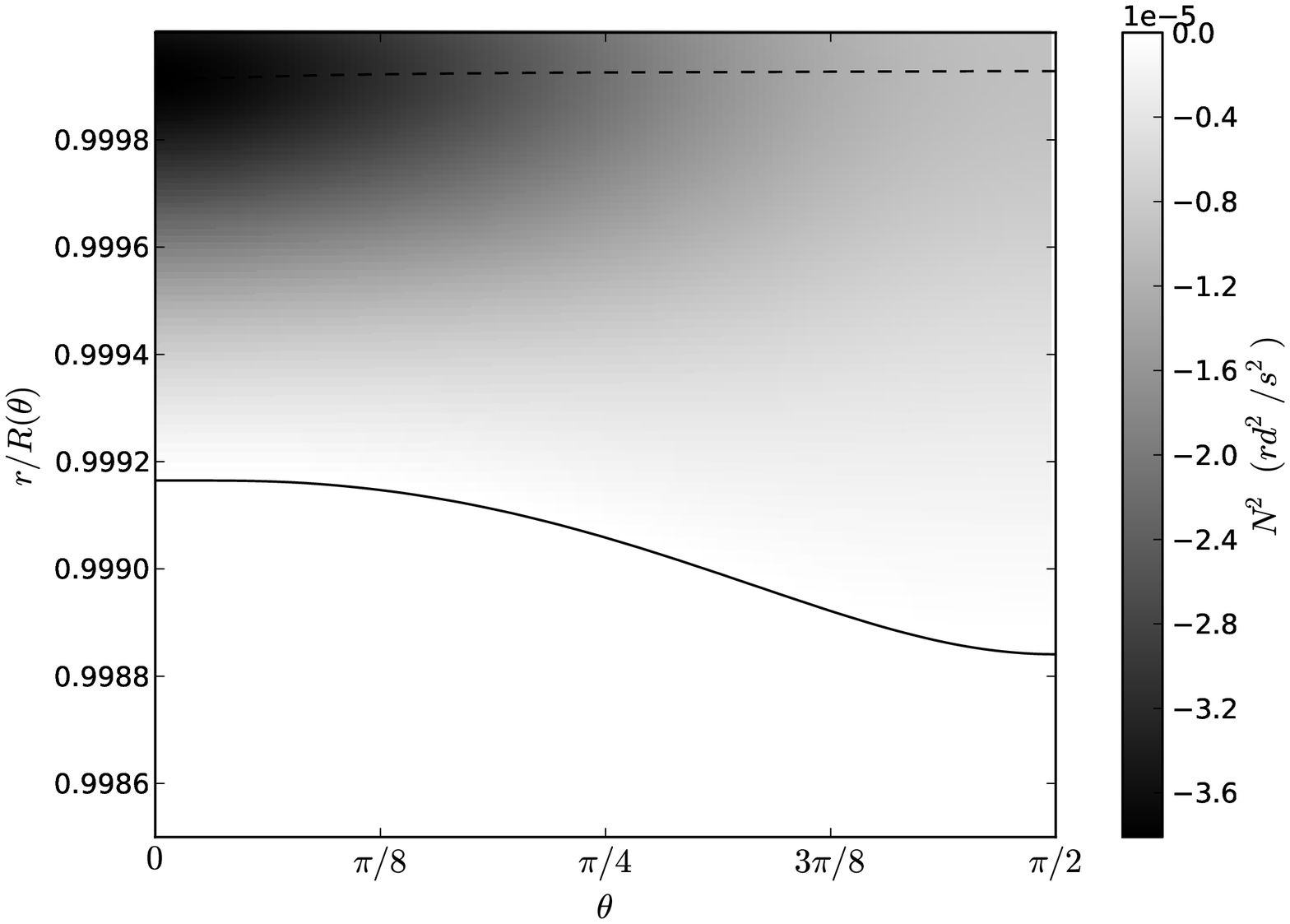}}
\caption{Convectively unstable layers in the outer layers of Vega's
model as shown by the squared \BVF $N^2$. Left: a general view showing
the regions where $N^2<0$. These are limited by the solid lines.
In the top panel, there are two convective layers, and the shallowest
one reaches the surface of the star. The dashed line marks the minimum
of $N^2$.  The layers associated with helium ionization region clearly
appear.  Left: Closer view of the surface layer. The grey scale
represents the values of $N^2$ in squared radians per second.}
\label{convec_vega}
\end{figure*}

\subsection{Comparison with observational data}

The most important test of the foregoing models naturally comes from a
comparison with observational data. The recent progress in optical and
infrared interferometry have led to the derivation of new precise measurements of
the fundamental parameters of some early-type stars of intermediate mass.

\begin{table}
\caption{Comparison between observationally derived
\cite[][]{monnier_etal10} parameters of the star $\alpha$ Oph and our
model. The mass fraction of hydrogen in the core is related to the age
of the star. P$_{\rm eq}$ and P$_{\rm pol}$ are the rotation periods in
days at equator and pole as predicted by the model. A solar composition
(X=0.7, Z=0.02) is assumed for the envelope.}

\begin{tabular}{lll}
\hline
                         & Observations     & Model \\
                         &                  &         \\
Mass  (M$_\odot$)        & $2.4^{+0.23}_{-0.37}$ & 2.22    \\
R$_{\rm eq}$ (R$_\odot$) & 2.858$\pm$0.015  & 2.865 \\
R$_{\rm pol}$ (R$_\odot$)& 2.388$\pm$0.013  & 2.385 \\
T$_{\rm eq}$ (K)         & 7570$\pm$124     & 7674 \\
T$_{\rm pol}$(K)         & 9384$\pm$154     & 9236 \\
L (L$_\odot$)             & 31.3$\pm$1       & 31.1 \\
V$_{\rm eq}$ (km/s)      & 240$\pm$12       & 242  \\
P$_{\rm eq}$ (days)      &                  & 0.598 \\
P$_{\rm pol}$ (days)      &                  & 0.616 \\
X$_\mathrm{core}$/X$_\mathrm{env.}$&        & 0.37 \\
\hline
\end{tabular}
\label{rasalhague}
\end{table}

\begin{table}
\caption{Same as in Table~\ref{rasalhague} but for Vega ($\alpha$
Lyr). Observationally derived values are from the concordance model of
\cite{monnier_etal12}. A sub-solar metallicity of X=0.7546, Y=0.2361,
and Z=0.0093 is adopted from \cite{yoon_etal08}.}
\begin{tabular}{lll}
\hline
                         & Observations     & Model \\
                         &                  &         \\
Mass  (M$_\odot$)        & 2.15$^{+0.10}_{-0.15}$    & 2.374   \\
R$_{\rm eq}$ (R$_\odot$) & 2.726$\pm$0.006 & 2.726 \\
R$_{\rm pol}$ (R$_\odot$)& 2.418$\pm$0.012 & 2.418 \\
T$_{\rm eq}$ (K)         & 8910$\pm$130    & 8973 \\
T$_{\rm pol}$(K)         & 10070$\pm$90    & 10070 \\
L (L$_\odot$)            & 47.2$\pm$2      & 48.0\\
V$_{\rm eq}$ (km/s)      & 197$\pm$23      & 205 \\
P$_{\rm eq}$ (days)      &                 & 0.672 \\
P$_{\rm pol}$ (days)     &                 & 0.697 \\
X$_\mathrm{core}$/X$_\mathrm{env.}$&       & 0.271\\
\hline
\end{tabular}
\label{vega}
\end{table}

\begin{table}
\caption[]{Same as in Table~\ref{rasalhague} but for $\alpha$
Leo. Observationally derived values are from \cite{che_etal11}.}
\begin{tabular}{lll}
\hline
                            &Observations     & Model \\
                            &                 &       \\
M (M$_\odot$)               & 4.15$\pm$0.06  & 4.10\\
R$_\mathrm{eq}$ (R$_\odot$) & 4.21$\pm$0.07  & 4.24\\
R$_\mathrm{pol}$ (R$_\odot$)& 3.22$\pm$0.05  & 3.23\\
T$_\mathrm{eq}$ (K)         & 11010$\pm$520  & 11175\\
T$_\mathrm{pol}$ (K)        & 14520$\pm$690  & 14567\\
L (L$_\odot$)               & 341$\pm$27     & 351\\
v$_\mathrm{eq}$ (km/s)      & 336$\pm$24     & 335\\
P$_{\rm eq}$ (days)      &                  & 0.641 \\
P$_{\rm pol}$ (days)      &                 & 0.658 \\
X$_\mathrm{core}$/X$_\mathrm{env.}$   &     & 0.5   \\
\hline
\end{tabular}
\label{regulus}
\end{table}

We have selected three such stars, namely $\alpha$ Lyr (M $\sim 2.2$
M$_\odot$), $\alpha$ Oph (M $\sim 2.2$ M$_\odot$), and $\alpha$ Leo (M
$\sim 4.1$ M$_\odot$). The star $\alpha$ Oph (Ras Alhague) has also been successfully
modelled by \cite{deupree_etal12}.

Data from these three stars comes from the interferometric observations
with CHARA
\cite[][]{Aufdenbergetal06,zhao_etal09,che_etal11,monnier_etal12}. In
addition, $\alpha$ Oph is the primary of binary system whose
orbit constrains its mass to the interval $2.03\leq M/M_\odot\leq
2.63$ \cite[][]{hinkley_etal11}. It has also been recently
identified as a $\delta$-Scuti star displaying 57 frequencies
\cite[][]{monnier_etal10}. All these data make Ras Alhague an interesting
test bed for models of rapidly rotating stars. Because Vega is a photometric
standard, it has been studied in detail for many years \cite[e.g.][and
references therein]{Aufdenbergetal06,takeda_etal08,monnier_etal12}.
Regulus is also an interesting case because it may owe its fast rotation
to the swallowing of the envelope of its now white dwarf companion
\cite[][]{rappaport_etal09}.

In Tables~\ref{rasalhague}, \ref{vega}, and \ref{regulus}, we compare our
models to these data. Three parameters have been adjusted: the mass,
the angular velocity, and the mass fraction of hydrogen in the core. The
last parameter allows us to take the hydrogen depletion into account 
in the core due to time evolution. This is our proxy for time evolution
on the main sequence. A solar metallicity is used for the envelope
(X=0.7 and Z=0.02), except for Vega where we use Z=0.0093 and X=0.7546
following \cite{yoon_etal08}.

The comparison shown by these tables is quite encouraging. We shall not
discuss the matching between data and models in more detail. Indeed, the
observed quantities are model dependent; actually, they have been derived
using a Roche model with solid body rotation and ad hoc gravity darkening
laws. A full discussion, which should include the spectral energy
distribution and the oscillation spectrum when relevant, would require
a dedicated work and is therefore beyond the scope of the present paper.

\subsection{Surface convection}

\begin{figure}
\resizebox{\hsize}{!}{\includegraphics{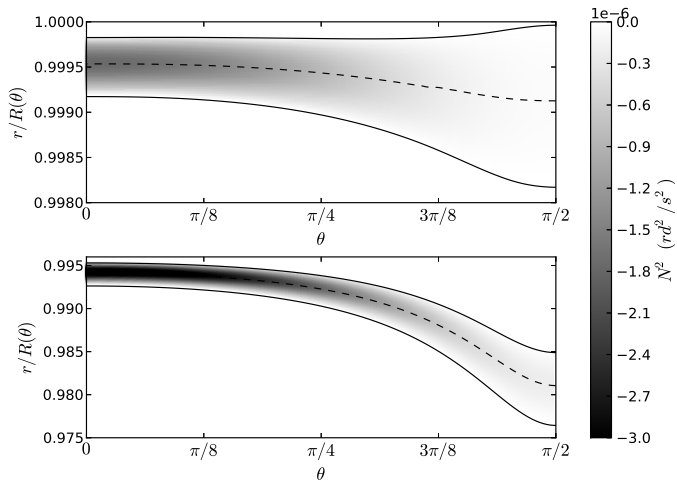}}
\caption{Layers with unstable stratification in the upper envelope of
the model for $\alpha$ Leo.}
\label{ioniz_regulus}
\end{figure}

While the most important convective region for stars with a mass above
two solar masses is the core, some residual convection may affect the
surface layers. In Fig.~\ref{convec_vega}, we display the regions where
the squared \BVF\ is negative for a Vega-like model. It shows that a
thin convective layer affects the surface due to the hydrogen ionization
peak. Two other layers associated with the first and second ionizations
of helium are also shown. As an effect of rotation, these layers thicken
and deepen in the equatorial region. In the more massive star
$\alpha$~Leo, the unstably stratified layers are below the surface
in the polar regions, but the equatorial temperature drop lets the
outermost layer
almost touch the surface (see Fig.~\ref{ioniz_regulus}). These regions,
if they actually develop some thermal convection, may participate in the
line broadening with micro- or macroturbulence.

\subsection{Differential rotation of intermediate-mass stars}

The essential progress that has been accomplished in these 2D-models
with respect to previous attempts is that differential rotation is no
longer arbitrary. It is the solution of fluid dynamics equations and is the
response to the actual baroclinic torque that exists in all rotating
stars. When stars rotate rapidly and do not lose angular momentum,
this torque is the main driver of the differential rotation. Reynolds
stresses from the convective core are also possibly influential, but
presently their functional form is unknown.

To have an idea of the resulting differential rotation in
intermediate-mass stars, we explore, in a rather systematic manner, the
dependence of differential rotation with general quantities that may
influence it. We therefore examine the dependence with mass, metallicity,
and angular velocity.

We first note that in all cases investigated, this differential rotation
is stable with respect to the centrifugal instability: the $z$-component
of the specific angular momentum always increases with the distance to
the rotation axis, except for a small jump at the CEI when the core is
evolved. This is a consequence of the weak differential
rotation generated by the baroclinic torques as shown below.

\begin{figure}
\resizebox{\hsize}{!}{\includegraphics{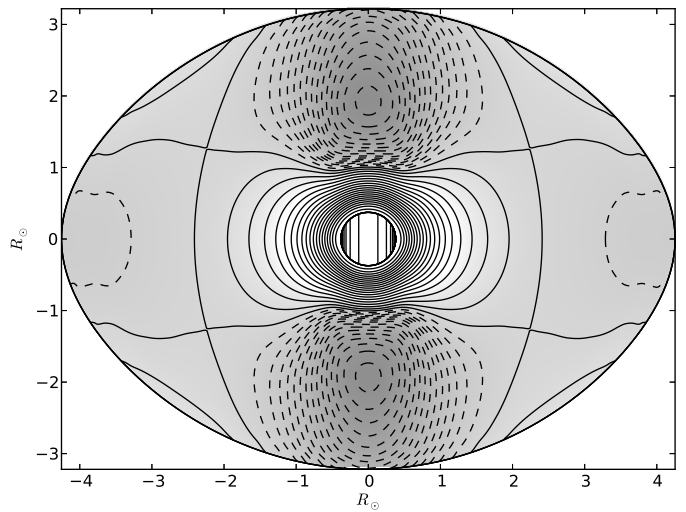}}
\caption{Differential rotation of Regulus. Dashed isocontours and dark
background represent slower regions, while clear or white background and
solid isocontours show faster regions.}
\label{diffrotreg}
\end{figure}

\begin{figure}
\resizebox{\hsize}{!}{\includegraphics{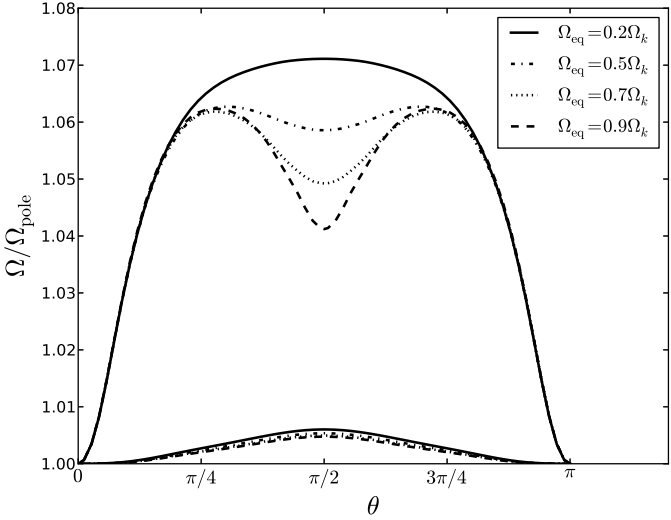}}
\caption{Latitudinal differential rotation on the surface of a 3
$M_\odot$ ZAMS star for several values of the equatorial rotation velocity
($\Omega_\mathrm{eq}$). The group of lower curves shows the latitudinal
variations of angular velocity at the core-envelope boundary.}
\label{fig:w_surf}
\end{figure}

\begin{figure}
\resizebox{\hsize}{!}{\includegraphics{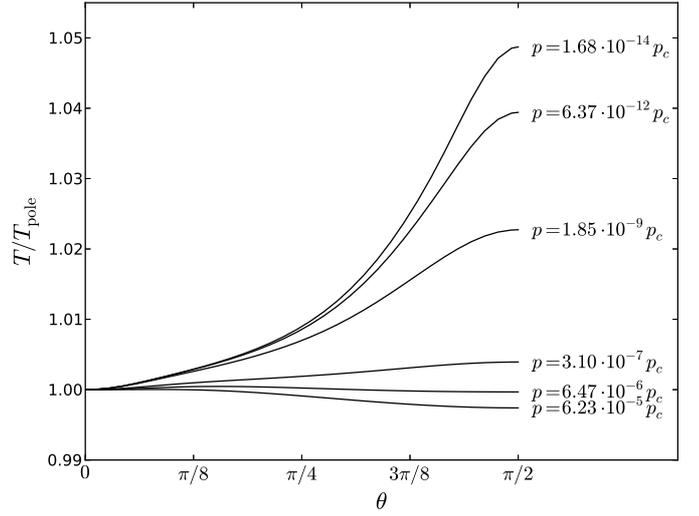}}
\caption{Latitudinal variations of temperature on an isobar at various
depth. The change from a decreasing function to an increasing one occurs
near the isobar that crosses the rotation axis at a fractional radius
of $r\sim0.6$.}
\label{tvar_isobar}
\end{figure}

\begin{figure}
\resizebox{\hsize}{!}{\includegraphics{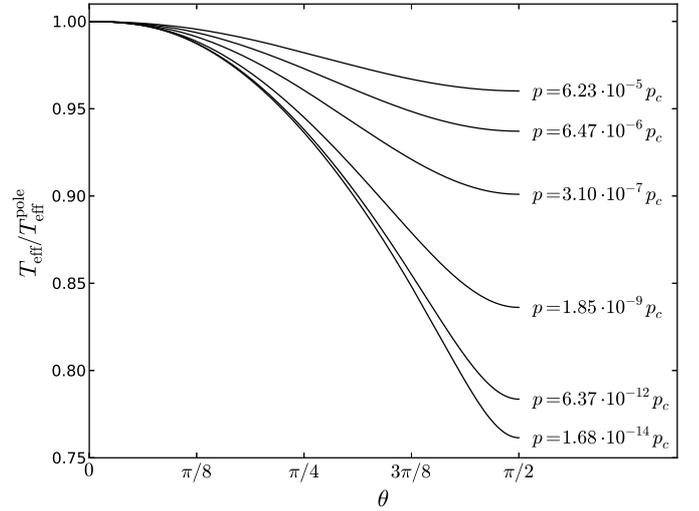}}
\caption{Latitudinal variations of effective temperature on an
isobar at various depth. Here, the effective temperature is defined as
\mbox{$T_\mathrm{eff}=\left(F^{(p)}/\sigma\right)^{1/4}$}, where $F^{(p)}$
is the energy flux crossing the isobar.}
\label{teff_isobar}
\end{figure}

\begin{figure}
\resizebox{\hsize}{!}{\includegraphics{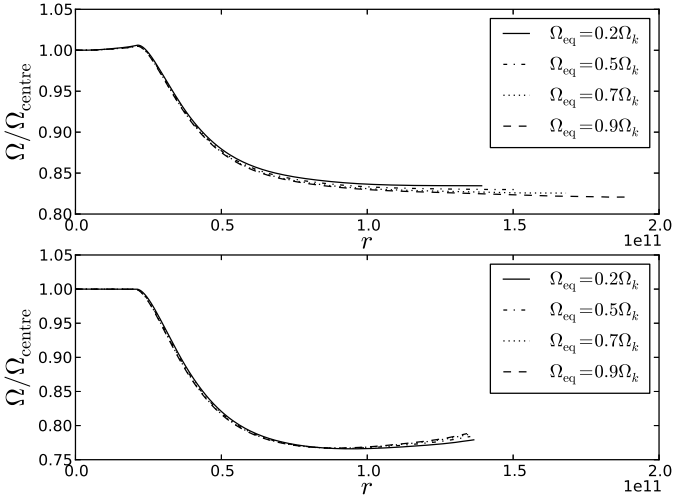}}
\caption{Radial profile of differential rotation at the equator (top)
and the pole (bottom) of a 3 $M_\odot$ star for several values of the
equatorial rotation velocity ($\Omega_\mathrm{eq}$). }
\label{fig:w_rad}
\end{figure}

A first general view of the differential rotation is given in
Fig.~\ref{diffrotreg} for the model of Regulus. As emphasized by the
surface latitudinal and radial variations on a ZAMS (homogeneous)
three solar mass star model, the differential rotation never exceeds 7~\%
horizontally and 25~\% radially (see Fig.~\ref{fig:w_surf} and
\ref{fig:w_rad}). As shown in Fig.~\ref{fig:w_surf} the differential
rotation is not necessarily a monotonous function of the latitude and, at
fast rotation rates, latitudes at $\pm40^\circ$ rotate slightly more rapidly (2~\%)
than the equator. This is at odds with our previous result
\cite[][]{ELR07}. The difference likely comes from the unrealistic
boundary conditions used in \cite{ELR07} where the star was confined in
a spherical container. The non-monotonous behaviour is only present in
the outer part of the envelope. It is related to the
change in the temperature variation with latitude on an isobar, since
these variations control the baroclinic torque \cite[][]{ELR07}. Indeed
as shown in Fig.~\ref{tvar_isobar} in the interior the temperature
decreases from pole to equator, while it increases near the surface. Such
a change affects the baroclinic torque and the ensuing velocity field. The
reason to this change is not clear.
The effective temperature representing the radiative flux,
as shown in Fig.~\ref{teff_isobar}, always decreases from pole to
equator, as expected.

The radial profiles displayed in Fig.~\ref{fig:w_rad} show the almost rigid
rotation of the core (as there is no baroclinic torque there). As noted
above, the radial variations are stronger than the latitudinal ones, but
they are essentially concentrated near the core (namely between $r_c$
and $2r_c$). Since the latitudinal differential rotation is much less, we
note that in this neighbourhood of the core, the differential rotation
is essentially shellular (see also Fig.~\ref{diffrotreg}). This remark
prompted us to compare the radial differential with that obtained from
the previous 1D-models, including rotation according to the approach of
\cite{zahn92}. Works like those of
\cite{Denissenkov_etal99} and \cite{meynet_maeder00} have shown that
an initial solid body rotation rapidly relaxes towards a quasi-steady
shellular differential rotation like the one displayed in
Fig.~\ref{fig:w_rad}. Our model shows that this shellular flow is just
the differential rotation driven by the baroclinic torque of the
configuration so does not depend on a prescription for the viscosity.
As a consequence the 1D and 2D models are in good agreement for the
amplitude of this differential rotation. For the 20~M$_\odot$-model of
\cite{meynet_maeder00}, we find a $\delta \Omega/\Omega\sim0.18$, while
their 1D-model gives $\delta \Omega/\Omega\sim0.21$.

\begin{figure}
\resizebox{\hsize}{!}{\includegraphics{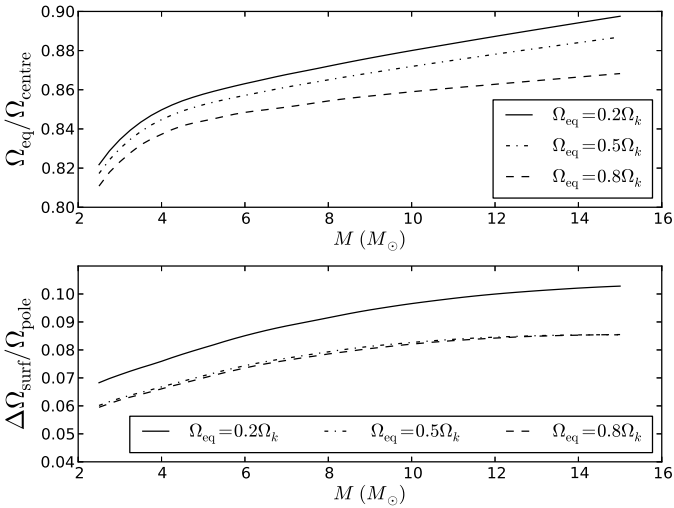}}
\caption{Top: Radial differential rotation, taken at equator, for ZAMS
(homogeneous) models as a function of mass. Bottom: surface
differential rotation for the same stars. The chemical composition is solar.}
\label{fig:w_mass}
\end{figure}

The dependence of differential rotation with respect to the mass of the
stars is illustrated in Fig.~\ref{fig:w_mass}. There we see that the
trend is for radial differential rotation to decrease with mass, while
latitudinal variations increase. This is a consequence of the variations
in the \BVF\ $\calN$ with mass. As mass increases, the peak of $\calN^2$
 near the core decreases, and therefore the baroclinic torque is weaker,
leading to a milder radial differential rotation.

\begin{figure}
\resizebox{\hsize}{!}{\includegraphics{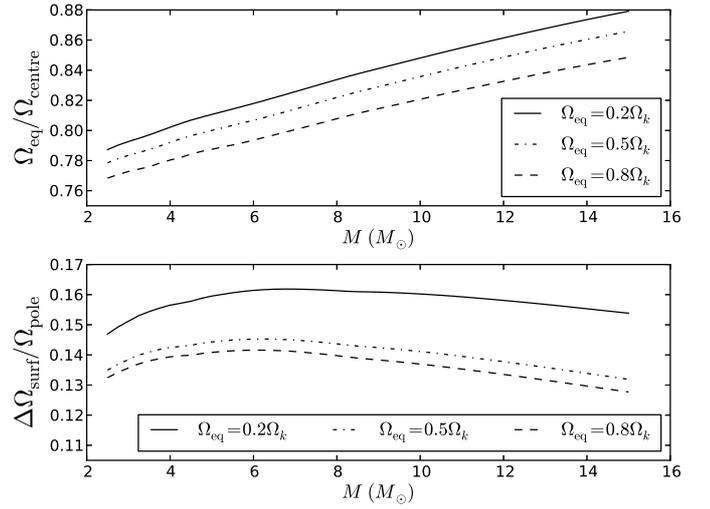}}
\caption{Same as Fig.~\ref{fig:w_mass} but stars with Z=0 metallicity.}
\label{fig:w_massZ0}
\end{figure}

Decreasing metallicity obviously increases differential rotation as
shown by Fig.~\ref{fig:w_massZ0}. The stronger latitudinal differential
rotation is conspicuous. We understand this effect as a consequence of
the reduction of the opacity when metals are removed. The gas is then
more efficient at heat transport, so temperature gradients are
steeper, making the baroclinic torque and the differential rotation
stronger.

\begin{figure}
\resizebox{\hsize}{!}{\includegraphics{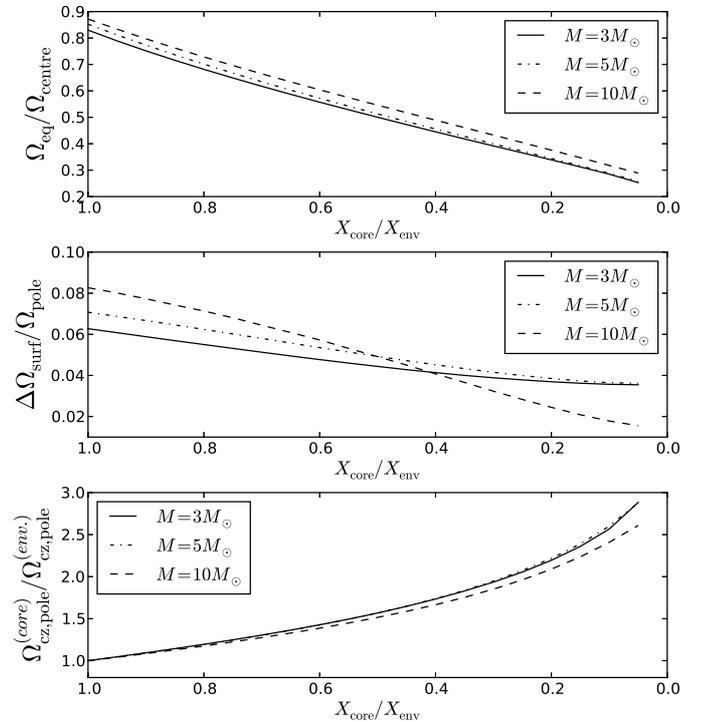}}
\caption{Top: Variation in the radial differential rotation with mass
fraction of hydrogen in the core. The value in the envelope is held
fixed at 0.7, while that of the core is decreased as would be the case
when the stars evolve. The angular velocity is half of the break-up:
$\Omega_\mathrm{eq}=0.5\Omega_k$. Middle: Same as top but for the
surface differential rotation. Bottom: The ratio of the angular
velocity at the pole of the core and at the bottom of the envelope on
the rotation axis.}
\label{fig:w_Xc}
\end{figure}

Finally, we tried to evaluate the effects of time evolution along the
main sequence. For early-type stars, the variations in the hydrogen
mass fraction in the convective core can serve as a proxy of this
evolution. Figure~\ref{fig:w_Xc} (top and middle) clearly shows that at a
given  fraction of the break-up angular velocity, hydrogen burning leads
to an increasing radial differential rotation and a decreasing latitudinal
variation.  This is related to the building up of a density jump at the
core-envelope interface as discussed above. As a consequence, the
angular velocity shows a rapid variation on this interface as shown in
Fig.~\ref{fig:w_Xc} (bottom).

\section{Conclusions}

We have presented the first realistic two-dimensional
models of a rotating star where the differential rotation and the
meridional circulation are calculated self-consistently with the stellar
structure. In particular, we have devised a method of computing
the velocity field without having to compute the Ekman boundary layers.
These layers are indeed extremely thin and would require too much spatial
resolution to be included in the solution. The account of the boundary
layer is, however, needed to lift the indetermination of the inviscid
solution with respect to geostrophic flows. The present solution therefore
includes then minimum viscosity for making the differential rotation
well-defined and insuring the balance of angular momentum flux between
meridional advection and diffusion. This simplification does not allow
the inclusion of a Stewartson layer, which requires the full viscosity
terms around the core.

The accuracy of the models have been certified through various tests:
\begin{itemize}
\item the spectral convergence,
\item the monitoring of round-off errors,
\item the virial test,
\item the energy balance test.
\end{itemize}

The first three tests usually show relative errors less than 10$^{-10}$,
while the energy balance yields much higher errors of the order of 10$^{-5}$
because of the rapid variations in opacity. Of course, the computation
of models close to break-up is less precise because of the development
of the equatorial cusp (making $\theta$-derivatives singular).

The resulting models have been compared to one-dimensional models
computed with well-tested codes like CESAM or TGEC in the non-rotating
case. Slight differences of the order of or less than a percent are
perceptible. They most probably come from our use of an analytic formula
for the nuclear heating instead of a nuclear network as is standard in
1D codes.

Finally, we compared our models to the observational data of some
rapidly rotating stars: namely $\alpha$ Lyr, $\alpha$ Oph, and $\alpha$
Leo. These stars have recently been studied in detail with interferometers
operating in the visible or infrared. Our models nicely match the
observationally derived parameters of these stars, suggesting that
they are not far from reality and confirming the previous good matching
of the gravity darkening found by \cite{ELR11}.

From these reassuring results we have studied the properties of
the differential rotation of baroclinic origin that pervade the
radiative envelope of intermediate-mass stars. The results show
that it remains small for a homogeneous star: a few percent in latitude
on the surface and less than 25~\% radially. Differential rotation
seems to decrease with increasing mass and metallicity. Besides, core
evolution increases the radial differential rotation because of the torque
coming from the density jump at the core-envelope interface. As shown in
Fig.~\ref{fig:w_Xc}, by the end of the main sequence, the core rotation
may reach three times that of the envelope. This result is important
in view of the interpretation of the core rotation of red giants stars
descending from intermediate-mass stars. This rotation is now available
from asteroseismology \cite[][]{mosser_etal12}.  Before the advent of a
full account of dynamical evolution in two-dimensional models, our results
suggest that baroclinic torques may lead to red giants with fast-rotating cores.

The foregoing two-dimensional models give a new view of the structure
and the dynamics of rapidly rotating stars, but they raise many new
questions as well. We have mentioned that of the temperature profile on isobars
that behaves non-monotonically and for which no simple explanation
was found, but the most challenging questions are certainly those around the
core-envelope interface including the Stewartson layer that inevitably
develops along the tangent cylinder of the convective core. Solutions
require the account of viscosity and the diffusion of elements. However,
Ekman layers still need to be avoided because they are much too thin
to be solved numerically. The complete solution likely necessitates a
combination of the asymptotic solution that we derived in this work and
a numerical solution of the full viscous operator in the interior of
the star. The density variations of the core-envelope boundary induced
by core evolution should also be smoothed by some diffusion effects;
nevertheless, they lead to a strong gradient of angular velocity that
emphasizes the mixing of the core and the base of the envelope.  One may
naturally wonder if this dynamical feature would lead to some faked core
overshooting that is confused with a true overshoot.

Finally, the next challenging issue in this modelling of rapidly
rotating stars will be including time evolution so as to monitor
the evolution of the distribution of angular momentum and determine the
consequences of the famous rotational mixing that is at the heart of
the specific chemical evolution of rotating stars \cite[][]{MM00}.

\begin{acknowledgements}
We are grateful to Sylvie Theado for providing us with 1D-models
computed with the Toulouse-Geneva Evolution Code, to Yveline
Lebreton for letting us use the one-dimensional CESAM code, and to
the referee Georges Meynet for his detailed comments on the first
version of our manuscript.
The authors acknowledge the support of the French Agence Nationale de
la Recherche (ANR), under grant ESTER (ANR-09-BLAN-0140).  This work
was also supported by the Centre National de la Recherche Scientifique
(CNRS, UMR 5277), through the Programme National de Physique Stellaire
(PNPS). The numerical calculations have been carried out on the CalMip
machine of the `Centre Interuniversitaire de Calcul de Toulouse' (CICT),
which is gratefully acknowledged.
\end{acknowledgements}

\bibliographystyle{aa}
\bibliography{/home/virgo/tex/biblio/bibnew}

\appendix

\section{Boundary layer equations}
\label{appendix}
\subsection{Differential operators using boundary layer coordinates}

The boundary layer coordinates ($\xi$, $\tau$, $\varphi$) defined in Sect. \ref{bl_coords} form
a set of orthogonal coordinates with scale factors

\begin{equation}
\begin{array}{l}
\displaystyle h_\xi=\varepsilon \\
\displaystyle h_\tau=\frac{R(\tau)}{\cos\delta(\tau)}\\
\displaystyle h_\varphi=R(\tau)\sin\tau\;.
\end{array}
\end{equation}

It is useful to define a new coordinate $\ts$ that depends on $\tau$ alone:
\begin{equation}
\ts=R(\tau)\sin(\tau)\;.
\end{equation}
The relation between $\ts$- and $\theta$-derivatives is given by
\begin{equation}
\frac{\partial}{\partial\ts}=\frac{\cos\delta}{R\cos(\tau-\delta)}\frac{\partial}{\partial\tau}\;,
\end{equation}
and the associated scale factor is
\begin{equation}
h_{\ts}=\frac{1}{\cos(\tau-\delta)}\;.
\end{equation}

Based on the scale factors, the common differential operators can be expressed as

\noindent Gradient:
\begin{equation}
\nabla \phi=\frac{1}{\varepsilon}\frac{\partial\phi}{\partial\xi}\vec{\hat\xi}
+\cos(\tau-\delta)\frac{\partial\phi}{\partial\ts}\vec{\hat\tau}
+\frac{1}{\ts}\frac{\partial\phi}{\partial\varphi}\vec{\hat\varphi}\;;
\end{equation}

\noindent Divergence:
\begin{equation}
\nabla\cdot\vec v=\frac{1}{\varepsilon}\frac{\partial v_\xi}{\partial\xi}
	+\frac{\cos(\tau-\delta)}{\ts}\frac{\partial}{\partial\ts}\left(\ts v_\tau\right)
	+\frac{1}{\ts}\frac{\partial v_\varphi}{\partial\varphi}\;;
\end{equation}

\noindent Curl:
\begin{equation}
\begin{array}{rl}
\nabla\times\vec v=&\displaystyle\left(\frac{\cos(\tau-\delta)}{\ts}\frac{\partial}{\partial\ts}(\ts v_\varphi)
	-\frac{1}{\ts}\frac{\partial v_\tau}{\partial\varphi}\right)\vec{\hat\xi}\\
&\displaystyle+\left(\frac{1}{\ts}\frac{\partial v_\xi}{\partial\varphi}
	-\frac{1}{\varepsilon}\frac{\partial v_\varphi}{\partial\xi}\right)\vec{\hat\tau}\\
&\displaystyle+\left(\frac{1}{\varepsilon}\frac{\partial v_\tau}{\partial\xi}
	-\cos(\tau-\delta)\frac{\partial v_\xi}{\partial\ts}\right)\vec{\hat\varphi}\;;
\end{array}
\end{equation}

\noindent Laplacian:
\begin{equation}
\Delta\phi=\frac{1}{\varepsilon^2}\frac{\partial^2\phi}{\partial\xi^2}+
\frac{\cos(\tau-\delta)}{\ts}\frac{\partial}{\partial\ts}
\left(\ts\cos(\tau-\delta)\frac{\partial\phi}{\partial\ts}\right)
+\frac{1}{\ts^2}\frac{\partial^2\phi}{\partial\varphi^2}\;;
\end{equation}

\noindent Material derivative:
\begin{equation}
\begin{array}{l}
(\vec u\cdot\nabla)\vec v=
\displaystyle
	\left[\frac{u_\xi}{\varepsilon}\frac{\partial v_\xi}{\partial\xi}
	+\cos(\tau-\delta)u_\tau\frac{\partial v_\xi}{\partial\ts}
	+\frac{u_\varphi}{\ts}\frac{\partial v_\xi}{\partial\varphi}
	\right]\vec{\hat\xi}\\
\qquad+\displaystyle
	\left[\frac{u_\xi}{\varepsilon}\frac{\partial v_\tau}{\partial\xi}
	+\cos(\tau-\delta)\left( u_\tau\frac{\partial v_\tau}{\partial\ts}
	-\frac{u_\varphi v_\varphi}{\ts}\right)
	+\frac{u_\varphi}{\ts}\frac{\partial v_\tau}{\partial\varphi}
	\right]\vec{\hat\tau}\\
\qquad+\displaystyle
	\left[\frac{u_\xi}{\varepsilon}\frac{\partial v_\varphi}{\partial\xi}
	+\cos(\tau-\delta)\left( u_\tau\frac{\partial v_\varphi}{\partial\ts}
	+\frac{u_\varphi v_\tau}{\ts}\right)
	+\frac{u_\varphi}{\ts}\frac{\partial v_\varphi}{\partial\varphi}
	\right]\vec{\hat\varphi}\;.
\end{array}
\end{equation}

\subsection{Vorticity equation}

Let us recall the general form of the vorticity equation

\begin{equation}
\na\times(\vo\times\vu)-s\frac{\partial\Omega^2}{\partial z}\vec{\hat\varphi}=
\frac{\na \rho\times\na p}{\rho^2}+E\nu\Delta\vo+E\vec{\calM}_\mu(\vu)\;.
\end{equation}
On the other hand, the interior inviscid solution satisfies the equation
\begin{equation}
-s\left(\frac{\partial\Omega^2}{\partial z}\right)_\mathrm{int.}\vec{\hat\varphi}=
\frac{\na \rho\times\na p}{\rho^2}\;.
\end{equation}
Subtracting the two equations, we have
\begin{equation}
\label{eq:app_vort1}
\na\times(\vo\times\vu)-2\Omega s(\vec{\hat z}\cdot\vec{\hat\xi})\beta\vec{\hat\varphi}=
E\nu\Delta\vo+E\vec{\calM}_\mu(\vu)\;,
\end{equation}
where we have used
\begin{equation}
\nabla\Omega=(\nabla\Omega)_\mathrm{int.}+\beta\vec{\hat\xi}\;.
\end{equation}

Within the boundary layer, the two components ($u_\xi$, $u_\tau$),
of the velocity field can be written as

\begin{equation}
u_\xi=\varepsilon^2 U+\vec{\hat\xi}\cdot\vec u_\mathrm{int}
\end{equation}
\begin{equation}
u_\tau=\varepsilon V+\vec{\hat\tau}\cdot\vec u_\mathrm{int}\;,
\end{equation}
where the interior solution $\vec u_\mathrm{int}$ is $\sim O(\varepsilon^2)$.
The associated vorticity is
\begin{equation}
\vec\omega=\nabla\times\vec u=\omega\vec{\hat\varphi}=\left(\frac{\partial V}{\partial\xi}
+O(\varepsilon)\right)\vec{\hat\varphi}\;.
\end{equation}
It can be shown that the first term in (\ref{eq:app_vort1}) is
\begin{equation}
\na\times(\vo\times\vu)\sim O(\varepsilon)\;,
\end{equation}
while the last term depends on second derivatives of the velocity field, so
\begin{equation}
E\vec{\calM}_\mu(\vu)\sim O\left(\frac{E||u||}{\varepsilon^2}\right) \sim O(\varepsilon)\;.
\end{equation}
However, the remaining term $E\nu\Delta\vo$ is $\sim O(1)$. Indeed,

\begin{equation}
\begin{array}{rl}
E\Delta\vec \omega&=\displaystyle E\Delta\left(\omega\vec{\hat\varphi}\right)
=E\left(\Delta\omega-\frac{\omega}{\ts^2}\right)\vec{\hat\varphi}\\
&\displaystyle =\left(\frac{\partial^3 V}{\partial\xi^3}+O(\varepsilon)\right)\vec{\hat\varphi}\;.
\end{array}
\end{equation}
Introducing this in (\ref{eq:app_vort1}) and rearranging terms, we
obtain at leading order in $\varepsilon$
\begin{equation}
2\Omega\ts\beta\cos(\tau-\delta)+\nu\frac{\partial^3 V}{\partial\xi^3}=0\;,
\end{equation}
where we have used the fact that, in the boundary layer, $\ts=s$ and 
$\vec{\hat z}\cdot\vec{\hat\xi}=\cos(\tau-\delta)$.

\subsection{Transport of angular momentum}

We start with the general equation in its compact form
\begin{equation}
\label{eq:app_lz}
\na\cdot{(\rho s^2\Omega\vu)}=E\na\cdot (\mu s^2\na\Omega)\;.
\end{equation}
Both terms in the equation are $\sim O(\varepsilon^2)$ for the interior
variables, but they are $\sim O(\varepsilon)$ within the boundary
layer. Let us start with the right-hand side

\begin{equation}
\begin{array}{rl}
E\na\cdot (\mu s^2\na\Omega)&=\varepsilon^2\na\cdot (\mu s^2(\na\Omega)_\mathrm{int.})+
\varepsilon^2\nabla\cdot(\mu s^2\beta\vec{\hat\xi})\\
&\displaystyle =\varepsilon\frac{\partial}{\partial\xi}(\mu s^2\beta)+O(\varepsilon^2)\;.
\end{array}
\end{equation}
The term $\na\cdot (\mu s^2(\na\Omega)_\mathrm{int.})$ only
depends on interior variables, so it is $\sim O(1)$. The
vertical derivative of an interior variable with respect to $\xi$ is
$\sim O(\varepsilon)$, so we can write

\begin{equation}
\label{eq:app_lz1}
E\na\cdot (\mu s^2\na\Omega)=\varepsilon\mu s^2\frac{\partial\beta}{\partial\xi}+O(\varepsilon^2)\;.
\end{equation}
Now, we calculate the remaining term
\begin{equation}
\na\cdot{(\rho s^2\Omega\vu)}=\rho s^2\vec u\cdot\nabla\Omega+ s^2\Omega\nabla\cdot(\rho\vec u)
+2s\Omega\rho\vec{\hat s}\cdot\vec u\;.
\end{equation}
From the continuity equation, we know that $\nabla\cdot(\rho\vec u)=0$, then
\begin{equation}
\na\cdot{(\rho s^2\Omega\vu)}=\rho s^2\vec u\cdot\nabla\Omega
+2s\Omega\rho\vec{\hat s}\cdot\vec u\;.
\end{equation}
Substituting $\nabla\Omega=(\nabla\Omega)_\mathrm{int}+\beta\vec{\hat\xi}$ and
$\vec u=\vec u_\mathrm{int}+\varepsilon^2U\vec{\hat \xi}+\varepsilon V\vec{\hat \tau}$, we have

\begin{equation}
\label{eq:app_lz2}
\na\cdot{(\rho s^2\Omega\vu)}=\varepsilon\rho s^2V\vec{\hat\tau}\cdot(\nabla\Omega)_\mathrm{int}
+2\varepsilon s\Omega\rho(\vec{\hat s}\cdot\vec{\hat \tau})V+O(\varepsilon^2)
\end{equation}
since $\vec u_\mathrm{int}\sim O(\varepsilon^2)$.  Substituting
(\ref{eq:app_lz2}) and (\ref{eq:app_lz1}) into (\ref{eq:app_lz}), up to
leading order in $\varepsilon$, we get

\begin{equation}
sV\vec{\hat\tau}\cdot(\nabla\Omega)_\mathrm{int}
+2\Omega(\vec{\hat s}\cdot\vec{\hat \tau})V=
\nu s\frac{\partial\beta}{\partial\xi}\;.
\end{equation}
We know that, in the boundary layer, $s=\ts$ and

\begin{equation}
\vec{\hat\tau}\cdot(\nabla\Omega)_\mathrm{int}=\cos(\tau-\delta)\frac{\partial\Omega}{\partial\ts}
\end{equation}
\begin{equation}
\vec{\hat s}\cdot\vec{\hat \tau}=\cos(\tau-\delta)\;;
\end{equation}
then, rearranging terms we obtain the final form

\begin{equation}
\lp2\Omega+\ts\frac{\partial\Omega}{\partial\ts}\rp V\cos(\tau-\delta)
-\ts\nu\frac{\partial\beta}{\partial\xi}=0\,.
\end{equation}

\end{document}